\documentclass[11pt,a4paper]{article}
\pdfoutput=1
\usepackage{jheppub}
\usepackage{rotating}
\usepackage{array}
\usepackage{amsmath}
\usepackage[normalem]{ulem}
\usepackage{slashed}
\usepackage{booktabs}
\usepackage[pdftex,table,dvipsnames]{xcolor}
\usepackage{units}
\usepackage{multirow}
\usepackage{multicol}
\usepackage{url}
\usepackage{graphicx}
\usepackage{xspace}
\usepackage{subfig}
\usepackage{color}
\usepackage{diagbox}
\usepackage{bbm}

\def \belletwo {Belle\,II\xspace}

\def \babar {BaBar\xspace}
\def \superkekb {SuperKEKB\xspace}
\def \bfactory {\textit{B}-factory\xspace}
\def \bfactories {\textit{B}-factories\xspace}

\preprint{DESY 20-226}

\title{
 Long-lived Dark Higgs and inelastic Dark Matter at \belletwo\\
}

\author[]{Michael Duerr,}
\author[1]{Torben Ferber,}
\author[1]{Camilo Garcia-Cely,}
\author[2 a,b]{Christopher Hearty,}
\author[1]{and Kai Schmidt-Hoberg}

\affiliation[1]{DESY, Notkestrasse 85, D-22607 Hamburg, Germany}
\affiliation[2]{ University of British Columbia, Vancouver, British Columbia, Canada V6T 1Z1$^a$; Institute of Particle Physics$^b$}

\emailAdd{michael.duerr@posteo.de}
\emailAdd{torben.ferber@desy.de}
\emailAdd{camilo.garcia.cely@desy.de}
\emailAdd{hearty@physics.ubc.ca}
\emailAdd{kai.schmidt.hoberg@desy.de}

\abstract{
Inelastic dark matter is an interesting scenario for light thermal dark matter which is fully consistent with all cosmological probes as well as direct and indirect dark matter detection. The required mass splitting between dark matter $\chi_1$ and its heavier twin~$\chi_2$ is naturally induced by a dark Higgs field which also provides a simple mechanism to give mass to the dark photon $A'$ present in the setup. The corresponding dark Higgs boson $h'$ is naturally the lightest dark sector state and therefore decays into Standard Model particles via Higgs mixing. In this work we study signatures with displaced vertices and missing momentum at \belletwo, arising from dark Higgs particles produced in association with dark matter. We find that \belletwo can be very sensitive to this scenario, in particular if a displaced vertex trigger is available in the near future.
}


\begin{document}

\maketitle

\flushbottom

\section{Introduction}

Thermal dark matter~(DM) is a well motivated and predictive scenario which can be probed with direct and indirect searches as well as with collider experiments. In recent years a lot of attention has focused on the MeV to GeV mass range as it is less constrained by the increasingly more stringent limits from direct searches~\cite{Aprile:2018dbl,Ren:2018gyx} while allowing for very interesting signatures in a number of current and future experiments~\cite{Batell:2009di,Batell:2009yf,Andreas:2012mt,Schmidt-Hoberg:2013hba,Essig:2013vha,Izaguirre:2013uxa,Morrissey:2014yma,Batell:2014mga,Dolan:2014ska,Krnjaic:2015mbs,Dolan:2017osp,Knapen:2017xzo,Beacham:2019nyx,Bernreuther:2019pfb,Bondarenko:2019vrb,Filimonova:2019tuy,Ballett:2019pyw,BelleII:2020fag,Bernal:2017mqb,Jodlowski:2019ycu,Baek:2020owl,Hostert:2020xku}. 

A simple scenario for light thermal DM that evades the strong bounds on residual DM annihilations from the Cosmic Microwave Background~(CMB) is the case where DM couples inelastically to Standard Model (SM) states~\cite{TuckerSmith:2001hy}.\footnote{Another option to circumvent these limits for $s$-wave DM annihilations is a resonantly enhanced cross-section at freeze-out~\cite{Bernreuther:2020koj}.} Here a sufficiently large mass splitting $\Delta \equiv m_{\chi_2}-m_{\chi_1}$ between the DM particle $\chi_1$ and its heavier twin $\chi_2$ ensures that {\it (i)} direct detection limits are basically absent and {\it (ii)} residual DM annihilations are no longer efficient during the time of the CMB. Limits from primordial nucleosynthesis~(BBN) still apply but are relevant only for masses $m_{\chi_1} \lesssim 10~\text{MeV}$~\cite{Depta:2019lbe}.
In the simplest setup inelastic DM is coupled to a massive dark gauge boson $A'$ which in turn kinetically mixes with the SM, a scenario that has been studied in a number of recent articles~\cite{Izaguirre:2015zva,Izaguirre:2017bqb,Berlin:2018jbm,Duerr:2019dmv}. These references were agnostic about the generation of the mass splitting (as well as the mass generation of the $A'$) but a natural setup to explain both would be a Higgs mechanism similar to what is realised within the SM.  Unitarity and perturbativity then suggest that the associated dark Higgs boson $h'$ cannot be much heavier than the gauge boson $A'$, while it can be significantly lighter~\cite{Kahlhoefer:2015bea,Duerr:2017uap,Darme:2018jmx}, implying that it is always present in the low-energy spectrum of the model and will therefore generally be very relevant for the resulting phenomenology.

In this article we study novel signatures associated with production and decay of the dark Higgs boson $h'$ at the \belletwo experiment.\footnote{A different signature at Belle involving a dark Higgs boson has been studied in~\cite{TheBelle:2015mwa}.} 
Given the strong bounds on the mixing angle with the SM Higgs boson together with the Yukawa-like coupling structure to SM states, the decay of the dark Higgs $h'$ will typically 
lead to displaced signatures. We point out that some regions of parameter space will not be covered with the current experimental configuration and that a displaced vertex trigger would be highly beneficial to
increase the sensitivity to this scenario. 

This article is structured as follows. Section~\ref{SEC:theory} is devoted to a description of the theoretical setup and the implications of this scenario for early universe cosmology. We also discuss constraints
that are complementary to the ones we explore in this paper.
In section~\ref{sec:belle} we describe in detail how the sensitivity of \belletwo to the displaced signatures is evaluated, while the results of this sensitivity study are presented in section~\ref{sec:results}.
In an appendix we provide further technical details regarding the inclusion of hadronic states in our analysis.

\section{Inelastic DM with a dark Higgs}
\label{SEC:theory}

\subsection{The model}
\label{SEC:the_model}

A stable Majorana fermion $\chi_1$ that can be excited to a state $\chi_2$ by absorbing a massive dark photon, $A'_\mu$, is usually called inelastic DM. The simplest realisation of this scenario consists of postulating a spontaneously broken $U(1)_X$ symmetry, under which all SM fields are singlets and a Dirac fermion, $\psi$, and a scalar, $\phi$, with charges 1 and 2, respectively. Before symmetry breaking, the Lagrangian describing the Dirac fermion reads
\begin{eqnarray}\label{eq:Lagrangian_psi}
\hspace{-20pt} \mathcal{L}_\psi &=& i \overline{\psi} \slashed{D} \psi - m_D \overline{\psi} \psi - f \phi  \overline{\psi^c} \psi+ \text{h.c.},
\end{eqnarray}
where we assume parity conservation for simplicity.\footnote{This implies that  $\psi_L$ and  $\psi_R$ couple to the scalar field $\phi$ in the same way.} Here $D_\mu =\partial_\mu- i g_X \hat{X}_\mu$ is the covariant derivative associated with the $U(1)_X$ symmetry, whose coupling constant is $g_X$.     After symmetry breaking, the scalar field acquires a vacuum expectation value (vev) $v_\phi$ and $\psi$ splits into two Majorana mass eigenstates. More precisely,
\begin{align}
\phi = \frac{v_\phi+\hat{h}'}{\sqrt{2}}\,, &&
\chi_1 =\frac{\psi-\psi^c}{\sqrt{2}}\,, &&\text{and}&&
\chi_2 =\frac{\psi+\psi^c}{\sqrt{2}}\,.
\end{align}
Note that we are working in the unitary gauge and correspondingly we do not write the Goldstone mode associated with $\phi$. In terms of these fields, the Lagrangian in Eq.~\eqref{eq:Lagrangian_psi} reads 
\begin{eqnarray}\label{eq:Lpsi}
\mathcal{L}_\psi &=&\frac{1}{2}\left( i \overline{\chi_1} \slashed{\partial} \chi_1 + i \overline{\chi_2} \slashed{\partial} \chi_2  - m_{\chi_1} \overline{\chi_1} \chi_1  - m_{\chi_2} \overline{\chi_2} \chi_2 \right)\\
&&+  \frac{i}{2} g_X \hat{X}_\mu (\overline{\chi_2} \gamma^\mu \chi_1 - \overline{\chi_1} \gamma^\mu \chi_2)
+ \frac{f}{2} \hat{h}' (\overline{\chi_1} \chi_1-\overline{\chi_2} \chi_2)\,, \nonumber
\end{eqnarray}
with
\begin{align}
\label{eq:chi2mass}
m_{\chi_2} = m_D+  f v_\phi && \text{and}&& m_{\chi_1} = m_D-  f v_\phi   \,.
\end{align}
The second line in Eq.~\eqref{eq:Lpsi} describes the inelastic interaction between DM $\chi_1$ and its excited state $\chi_2$ as well as the DM interactions with the neutral scalar $\hat{h}'$. 

In general a mixing term between the dark scalar field $\phi$ and the SM Higgs field $H$ is present in the Lagrangian, leading to a mixing of the flavour eigenstates $\hat{h}'$ and $\hat{h}$,
as determined by the overall scalar potential
\begin{eqnarray}\label{eq:potential}
\hspace{-10pt}
V(\phi,H)  =\lambda_{H} \left(H^{\dagger}H-\frac{v_H^2}{2}\right)^2
\hspace{-5pt}
+\lambda_{\phi}\left(\phi^{*}\phi -\frac{v_\phi^2}{2}\right)^2
\hspace{-5pt}
+
	\lambda_{\phi H}\, \left(H^{\dagger}H  -\frac{v_H^2}{2}\right)\left( \phi^{*}\phi-\frac{v_\phi^2}{2}\right).
\end{eqnarray}
Here write the SM scalar doublet as $H=\left(0, (v_H+\hat{h})/\sqrt{2}\right)^T$. In terms of the SM scalar and the dark Higgs, the scalar fields before diagonalisation (denoted by hats) read
\begin{equation}
\begin{pmatrix}  \hat{h} \\\hat{h}'\end{pmatrix} =
\begin{pmatrix}
c_\theta && s_\theta \\
-s_\theta && c_\theta
\end{pmatrix}
\begin{pmatrix}
h \\ h'
\end{pmatrix}\,.
\end{equation}
Likewise, the quartic couplings can be expressed in terms of the mixing angle, the vevs and the masses
\begin{align}\label{eq:quartic}
\lambda_H  = \frac{m_h^2 c^2_\theta + m_{h'}^2 s^2_\theta}{2 v_H^2}\,,&&\lambda_\phi  = \frac{m_h^2 s^2_\theta + m_{h'}^2 c^2_\theta}{2 v_\phi^2}\,&&\text{and}&&
 \lambda_{\phi H}=\frac{(m_{h'}^{2}-m_{h}^{2})\,s_{2\theta}}{2\,v_{H}\,v_{\phi}}\,.
\end{align}

Coming to the gauge sector of the theory, the most general Lagrangian includes a kinetic mixing term between the dark $U(1)_X$ and $U(1)_Y$ and is given by
\begin{align}
\label{eq:LGauge}
{ \mathcal{L}} = \mathcal{L}_\text{SM} - \frac{1}{4} \hat{X}_{\mu\nu} \hat{X}^{\mu\nu} -\frac{\epsilon}{2 c_\text{W}}  \hat{X}_{\mu\nu} \hat{B}^{\mu\nu} &&\text{with}&&  \mathcal{L}_\text{SM} \supset - \frac{1}{4} \left(\hat{B}_{\mu\nu} \hat{B}^{\mu\nu} +  \hat{W}^a_{\mu\nu} \hat{W}^{a\mu\nu}\right)\,,
\end{align}
together with additional terms from the covariant derivatives of the scalar Lagrangian that give mass to the gauge bosons. 
This is the (dark) Higgs mechanism, which also demands that the cubic interaction between one scalar and two gauge bosons must be proportional to the corresponding mass.  Since they are crucial for our work, we write them explicitly 

\begin{eqnarray}
\mathcal{L}_{\phi} &=& |D_\mu H|^2+|D_\mu\phi|^2 - V(\phi, H)\nonumber \\
&\supset&  \frac{1}{2} m_{\hat{Z}}^2 \left(1+ \frac{2\hat{h}}{v_H} \right) \hat{Z}_{\mu} \hat{Z}^{\mu}+ \frac{1}{2} m_{\hat{X}}^2 \left(1+ \frac{2\hat{h}'}{v_\phi} \right)\hat{X}_{\mu} \hat{X}^{\mu}\,. 
\label{eq:HiggsMechanism}
\end{eqnarray}  
We denote the gauge fields and the corresponding masses in the original basis before diagonalisation by hats, such that $\hat{B}_{\mu\nu}$, $\hat{W}_{\mu\nu}$, and $\hat{X}_{\mu\nu}$ are the field strength tensors of $U(1)_Y$, $SU(2)_L$, and $U(1)_X$, respectively. 
The gauge-boson diagonalisation has been comprehensively discussed in~\cite{Babu:1997st,Frandsen:2011cg}. Here we just emphasise the most relevant aspects for our work and refer the reader to that study for further details. The hatted fields $\hat{B}$, $\hat{W}$, and $\hat{X}$ are diagonalised and canonically normalised to 
obtain the physical $Z$-boson, the photon and the physical dark photon, $A^\prime_\mu$.  
Eq.~\eqref{eq:LGauge} implies that 
for sufficiently small masses $m_{A'}$ as studied in this article, the field $A'_\mu$ inherits the coupling structure of the photon to the SM fermions up to a common factor $\epsilon$. Moreover, the part of Eq.~\eqref{eq:HiggsMechanism} of interest in this work reads ${\cal L}_\phi\supset (m_{A'}^2/2) (1+2 \hat{h}'/v_\phi) A'_\mu A'^\mu$ plus small corrections of order ${\cal O}(\epsilon^2)$.

Overall the model contains two independent portals between the dark and visible sector, leading to a `two mediator' model with a rather complex phenomenology (see e.g.~\cite{Duerr:2016tmh,Darme:2017glc} for a recent discussion).

\subsection{Parameters of the model}
 As a consequence of the diagonalisation process, the mass parameters $m_{\hat{X}}$ and  $m_{\hat{Z}}$ in Eq.~\eqref{eq:HiggsMechanism} can be exchanged for the physical masses $m_{A'}$ and $m_{Z}$. Furthermore, the dark vev is determined by the dark photon mass
\begin{equation}
\label{eq:massAp}
v_\phi  = \frac{m_{\hat{X}}}{2 g_X} =   \frac{m_{A'}}{2 g_X} \left(1+\mathcal{O}(\epsilon^2)\right)\,,
\end{equation}
where the expansion assumes $m_{A'}<m_Z$. 
Taking into account the correlations according to Eqs.~\eqref{eq:chi2mass}, \eqref{eq:quartic} and \eqref{eq:massAp}, the dark sector has seven free parameters. Two of them characterise the dark photon: its mass,  $m_{A'}$, and the kinetic mixing, $\epsilon$. Likewise, $m_{h'}$ and $\theta$ specify the properties of the  dark Higgs. Finally, three parameters describe the DM: its mass, $m_{\chi_1}$, as well as its couplings to the dark photon, $g_X$, and to the dark Higgs, $f$. Note that the latter can be exchanged for the mass of the excited state by means of Eqs.~\eqref{eq:chi2mass} and \eqref{eq:massAp}. Finally, we introduce $ \alpha_D =g_X^2/4\pi$ and $\alpha_f = f^2/4\pi$  for convenience.

When using these parameters it is important to realise that not all combinations correspond to the perturbative regime. In particular, requiring that all couplings remain smaller than $\sqrt{4 \pi}$ directly implies that the dark Higgs $h'$ cannot be much heavier than the dark photon $A'$ for the parameters we are interested in. Explicitly, Eq.~(\ref{eq:massAp}) together with Eq.~(\ref{eq:quartic}) gives
\begin{equation}
m_{h'}^2 \lesssim \frac{1}{4\sqrt{\pi}\alpha_D}m_{A'}^2 
\end{equation}
assuming $\epsilon$ and $\theta$ are small and $\lambda_\phi < \sqrt{4\pi}$. 
We will indicate the corresponding non-perturbative region which violates this condition in the plots below.

\subsection{Dark matter} Due to the charge assignments described above, the $U(1)_X$ symmetry spontaneously breaks into a remnant global $Z_2$ group, under which  $\chi_1$ and $\chi_2$  are odd while  all SM fields, $h'$ and $A'$ are even. Such a symmetry is crucial for inelastic DM because it guarantees the absolute stability of our DM candidate, $\chi_1$. In contrast, the excited state may decay. For the mass splittings $\Delta= m_{\chi_2}-m_{\chi_1}$ of interest in this work, $\chi_2$ decays into $\chi_1$ plus a pair of leptons or even hadrons. In the former case, the decay rate is
\begin{equation}
     \Gamma_{\chi_2 \to \chi_1 l^+l^-}=
     \alpha_\text{em} \alpha_D \epsilon^2  \int^{\Delta^2}_{4m_l^2}ds  
    \frac{| \vec{p}_{\chi_1}|   (s-\Delta^2)\left(2s +(2m_{\chi_1}+\Delta)^2\right)(s+2m_l^2)(s-4m_{l}^2)^{1/2}}{ 6\pi m_{\chi_2}^2 s^{3/2} \left(s-m_{A'}^2\right)^2}  
        \,,
        \label{eq:chi2decayrate}
\end{equation}
where $|\vec{p}_{\chi_1}|$ is the momentum of $\chi_1$ in the  rest frame of $\chi_2$ (see Eq.~\eqref{eq:pchi1}) and $s$ is the invariant mass of the lepton pair. The hadronic decay rate can be calculated with a similar expression by setting $m_l = m_\mu$ and adding in the integrand the  experimentally obtained factor  $R(s) \equiv \sigma(e^+e^-\to\text{hadrons})/\sigma(e^+e^- \to \mu^+ \mu^-)$~\cite{Tanabashi:2018oca}. A detailed derivation of this expression is given in the appendix and improves on the approximation for the decay width $ \Gamma_{\chi_2}$ used in~\cite{Duerr:2019dmv}.

For the couplings of interest in this work, we find that  $\chi_2$ is unstable on cosmological scales and does not contribute to the DM abundance today. 
For a sufficiently light dark Higgs, bound states of $\chi_1$ or $\chi_2$ might also contribute to the DM abundance. They form at low velocities via the radiative emission of a dark Higgs due to the attractive Yukawa potential induced by $h'$: 
$\alpha_f\,e^{-m_{h'}r}/r$ (see e.g.~\cite{Ko:2019wxq}). As explained below, in this work we will focus on parameter regions where such a process is kinematically closed when $\chi_1$ is non-relativistic. More precisely, we will assume that the corresponding binding energy is smaller than $m_{h'}$. This implies that only $\chi_1$ contributes to the DM density today. 

\subsection{Cosmology}

Before we discuss possible signatures at colliders in detail let us briefly describe the cosmological evolution of our scenario and delineate the interesting regions in parameter space.

\subsubsection{DM abundance} 
To have any hope of producing dark sector particles at Belle II, the portal interactions should not be too small to allow for a sizeable production cross section.
In fact, the corresponding required values for $\epsilon$ or $\theta$ are typically large enough to bring DM into 
thermal equilibrium with SM particles in the Early Universe.\footnote{In principle rather small mixing angles of the dark Higgs can be constrained by \belletwo, which are insufficient to keep up the thermal equilibrium between the dark and visible sectors until DM freeze-out. For this region in parameter space the calculation of the DM abundance is more involved~\cite{Bringmann:2020mgx}. For the signature we are interested in, however, a sizeable value of $\epsilon$
will always guarantee thermal equilibrium and applicability of the standard thermal freeze-out prescription.} 
The initial DM abundance was thus very large and we must allow for a process reducing it to its current value or below. 
One possible process is DM annihilation which must have a velocity-averaged annihilation cross section, $\langle\sigma v\rangle$,  greater than roughly $\unit[10^{-26}]{cm^3/s}$ at early times in order to sufficiently reduce the abundance under standard cosmological assumptions. 
While smaller annihilation cross-sections may be viable for non-standard cosmological histories (e.g.\ featuring an early period of vacuum or matter domination),
we will assume standard cosmology in this work.
To evaluate the DM relic abundance we implement our inelastic DM model within \texttt{micrOMEGAs}~\cite{Belanger:2018ccd}, which calculates all the cross sections for the DM (co-)annihilation processes (see Table \ref{table:process}) to obtain the DM abundance today.\footnote{As \texttt{micrOMEGAs} does not account for hadronisation and naively calculates the annihilation cross section into light quarks, we modify these annihilation channels by hand making use of the experimentally inferred ratio $R(s)$ as described in \cite{Duerr:2019dmv}. For most of our parameter space this turns out to be completely irrelevant however as the annihilation cross section is dominated by $\chi_1 \chi_1 \to h' h'$.} 
In Table~\ref{table:process} we also indicate the orbital angular momentum of the annihilation channels.  In the case of $s$-wave annihilations, the corresponding cross sections $\langle \sigma v\rangle$ are independent of the velocity and do not change as the Universe evolves. 
In contrast, for $p$-wave processes, the cross section scales like $\langle \sigma v\rangle \propto v^2$ at leading order in $v$ and the annihilation rates are substantially reduced when DM becomes non-relativistic at late times, for instance during the formation of the CMB and thereafter.

\begin{table}[]
    \centering
    \begin{tabular}{c|c|c|c}
\multirow{3}{*}{     Annihilation }& 
\multirow{3}{*}{ Type of} &
\multirow{3}{*}{ Representative}&
\multirow{3}{*}{Relevant}
\\
&&&\\
channel & process & diagrams & couplings
\\\hline
 \multirow{5}{*}{  $\chi_1\chi_1 \to A'A'$}
       & 
\multirow{5}{*}{        $s$-wave} &
\multirow{5}{*}{ \includegraphics[ width=0.3\textwidth]{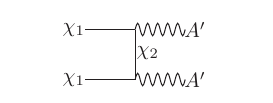}} &
\multirow{5}{*}{ $\alpha_D^2 $}
\\
&&&\\
&&&\\
&&&\\
\multirow{5}{*}{         $\chi_1\chi_2 \to A' \to \text{SM}\,\text{SM}$} & 
\multirow{5}{*}{ $s$-wave} &   \multirow{5}{*}{ \includegraphics[width=0.3\textwidth]{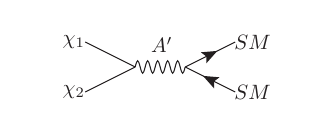}}& 
\multirow{5}{*}{ $\alpha_D  \alpha_\text{em}\,\epsilon^2$}\\
&&&\\
&&&\\
&&&\\
\multirow{5}{*}{        $\chi_1\chi_1 \to h'(\chi_1\chi_1)$} & 
\multirow{5}{*}{$s$-wave} &
\multirow{5}{*}{\includegraphics[width=0.3\textwidth]{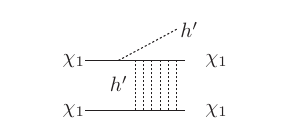}}& 
\multirow{5}{*}{$\alpha_f^4$}\\
&&&\\
&&&\\
&&&\\
\multirow{5}{*}{        $\chi_1\chi_1 \to h' \to \text{SM}\,\text{SM}$} &
\multirow{5}{*}{$s$-wave } &
\multirow{5}{*}{\includegraphics[width=0.3\textwidth]{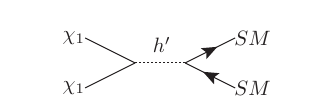}}& 
\multirow{5}{*}{$\alpha_f \,y_\text{SM}^2\theta^2$ }\\
&&&\\
&&&\\
&&&\\
\multirow{5}{*}{        $\chi_1\chi_1 \to h'h'$ }&
\multirow{5}{*}{$p$-wave}  & 
\multirow{5}{*}{\includegraphics[width=0.3\textwidth]{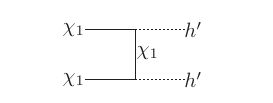}}&
\multirow{5}{*}{$\alpha_f^2$} \\
&&&\\
&&&\\
&&&\\
    \end{tabular}
    \caption{(Co-)annihilation channels involving DM. Here $(\chi_1\chi_1)$ and $\text{SM}$ respectively stand for DM bound state and Standard Model particle. $\alpha_\text{em}$ is the fine-structure constant and $y_\text{SM}$ the corresponding SM Yukawa coupling.  }
    \label{table:process}
\end{table}

\subsubsection{CMB constraints } All annihilation products of Table \ref{table:process} eventually decay into charged particles or photons. Consequently, DM annihilations in the period between recombination and reionisation inject energy into the CMB, potentially modifying its black-body shape or inducing  non-primordial anisotropies. The precise observations of the CMB by the Planck satellite
constrains the DM mass to be $m_\chi \gtrsim \unit[10]{GeV}$ for a thermal dark matter annihilation cross section at the time of the CMB~\cite{Ade:2015xua}.
This implies that for light DM to be viable, the annihilation cross section at late times needs to be suppressed compared to its value at DM freeze-out.
Inspecting the various possible annihilation channels in Table~\ref{table:process} we observe that all constraints can be evaded if we require
\begin{align}
\frac{1}{4}{\alpha_f^2} m_{\chi_1} <m_{h'}\lesssim m_{\chi_1} < m_{A'} \,.
\end{align}
In this case, the annihilation channels $\chi_1\chi_1 \to A'A'$ and  $\chi_1\chi_1 \to h'(\chi_1\chi_1)$~\cite{An:2016kie} will be kinematically closed at low velocities.
As $\chi_1\chi_1 \to \text{SM}\,\text{SM}$ is always negligible due to the smallness of the relevant couplings the overall annihilation cross section will naturally be dominated by either  $\chi_1\chi_2 \to A' \to \text{SM}\,\text{SM}$ or $\chi_1\chi_1 \to h'h'$ which are both suppressed at late times, either via the mass splitting $\Delta$ (leading to a suppressed abundance of $\chi_2$) or because of the $p$-wave nature of the annihilation. 
 As we will largely concentrate on parameter regions in which the mass splitting $\Delta$ is sizeable, the DM relic density will be dominantly set by the process $\chi_1\chi_1 \to h' h'$. Note that the inequality $m_{h'}\lesssim m_{\chi_1}$ is not strict: DM can still annihilate into a pair of slightly heavier dark Higgses due to thermal effects in the Early Universe~\cite{DAgnolo:2015ujb}. These are the so-called forbidden annihilation channels, which imply $m_{h'}\simeq m_{\chi_1}$ as well as sizeable values\footnote{See Ref.~\cite{Bernal:2017mqb}  for another production mechanism of inelastic DM leading to large couplings.} for $\alpha_f$ and thus relatively large values for $\Delta=m_{\chi_2}-m_{\chi_1}$. As we calculate the relic abundance using \texttt{micrOMEGAs}, these forbidden channels are automatically taken into account.

\subsection{Established constraints and future prospects}
Before we discuss the sensitivity of \belletwo to signatures including the dark Higgs in detail, let us briefly comment on complementary limits on this setup. Given that our model features two independent portal interactions, there are a variety of different searches that are potentially sensitive. In particular there are a number of searches which constrain either the vector or the Higgs portal individually. Starting with the vector portal, there are well known constraints from electroweak precision observables which apply independently of the specific couplings of the $A'$ to dark sector states, constraining $\epsilon \lesssim 3 \times 10^{-2}$ for dark photon masses below the $Z$ mass~\cite{Hook:2010tw}. In addition there are constraints from HERA measurements~\cite{Kribs:2020vyk}, which are slightly stronger for small $m_{A'}$. The latter constraint is expected to improve by an order of magnitude at the LHeC~\cite{Kribs:2020vyk}.
Generic dark photon searches on the other hand typically do not apply, as the $A'$ decays neither fully visibly nor fully invisibly in our scenario. 
This is different for the dark Higgs $h'$, which decays fully visibly as it is the lightest dark sector state in the regions of parameter space we consider, implying that searches for a Higgs-mixed scalar directly apply.
An updated compilation of current constraints can e.g.\ be found in \cite{Winkler:2018qyg}.

In addition to the general signatures above, there are a variety of experimental probes which constrain more specific signatures of inelastic dark matter.
Of potential relevance here are electron and proton beam dumps, \bfactories and direct detection experiments.
While dark matter direct detection at tree-level is kinematically impossible via $A'$ exchange for the mass splittings $\Delta$ we consider, loop-induced elastic scattering is generally present.
As discussed in~\cite{Duerr:2019dmv} however the elastic scattering cross section due to two dark photon exchanges is very suppressed.
On the other hand the dark Higgs $h'$ has diagonal couplings to $\chi_1$ which induces spin-independent scatterings with nuclei. As we consider rather small DM masses $m_{\chi_1}$ and mixing angles $\theta$ the resulting constraints, while potentially relevant for small dark Higgs masses, turn out to not be overly constraining. 

Another potential constraint comes from the requirement of successful primordial nucleosynthesis (BBN), which will in general be relevant for sufficiently light or sufficiently long-lived particles, see e.g.~\cite{Berger:2016vxi,Fradette:2018hhl,Depta:2020zbh}.
The dark Higgs $h'$ in particular can be very long-lived due to the extra Yukawa suppression of its couplings to light SM states. Note however that our setup is rather different from e.g.~\cite{Fradette:2018hhl} where 
\textit{only} a scalar mixing with the Higgs is studied and the corresponding constraints can therefore not be directly applied. In particular our setup will naturally imply thermalisation between the dark and visible sector even for small values of $\theta$ due to the sizeable values of $\epsilon$ we consider. While a dedicated study of BBN constraints on the current scenario is beyond the scope of this work, we don't expect
any impact on the parameter regions covered by \belletwo.

In addition, particles with mass $m \lesssim \unit[200]{MeV}$ can be copiously produced in the hot cores of supernovae and will lead to a new energy loss mechanism if they interact sufficiently weakly to escape. 
As we assume the dark Higgs $h'$ to be the lightest dark sector state and therefore to decay visibly, the limits from SN1987A on Higgs-mixed dark scalars should to a good approximation apply, see e.g.~\cite{Winkler:2018qyg}.
Nevertheless there are large intrinsic uncertainties associated to these limits and indeed the explosion mechanism of SN1987A has not been fully settled, potentially invalidating the bounds on light dark sector particles completely~\cite{Bar:2019ifz}.

Beam dumps are however potentially sensitive to the production of DM with subsequent scattering (or decay) in a far detector and relevant bounds come from various experiments, including LSND~\cite{deNiverville:2011it}, E137~\cite{Bjorken:1988as,Berlin:2018pwi}, MiniBoonNE~\cite{Aguilar-Arevalo:2017mqx} and NA64~\cite{NA64:2019imj}. 
As these experiments are sensitive in particular to small dark photon masses, $m_{A'} \ll 1\,\mathrm{GeV}$, they are complementary to the searches at \belletwo that we will discuss below. 

Coming to \bfactories such as \babar or \belletwo, a largely model-independent signature is the final state consisting of only a single photon (so-called mono-photon searches)
which naturally applies if a photon and a dark photon are produced in association and the $A'$ decays invisibly, i.e.\ $e^+ e^- \rightarrow \gamma A^\prime$, $A^\prime \rightarrow \text{invisible}$. 
It will however also apply to those regions of parameter space where the decay products of the $A'$ are visible but sufficiently long-lived so that they decay outside the detector.
In the next section we will reinterpret the BaBar mono-photon limit~\cite{Lees:2017lec} following the discussion in~\cite{Duerr:2019dmv} as well as give 
an overview of other possible signatures within \belletwo.

Finally there are also a large number of proposed future experiments, see e.g.\ figure~7 of Ref.~\cite{Berlin:2018jbm} for a comprehensive overview of limits on inelastic dark matter, including potential add-ons to the LHC such as FASER~\cite{Feng:2017uoz}, MATHUSLA~\cite{Chou:2016lxi}, and CODEX-b~\cite{Gligorov:2017nwh} or possible future beam dumps such as LDMX~\cite{Akesson:2018vlm} and SeaQuest~\cite{Berlin:2018pwi}. 
Also the bounds on the direct production and observation of the dark Higgs $h'$ will become ever more stringent, see e.g.\ \cite{Beacham:2019nyx} for a recent overview.

\section{Light dark Higgs and inelastic DM at \belletwo}
\label{sec:belle}

The current scenario can lead to a number of different signatures at \belletwo. One signature arises from direct production of the dark Higgs $h'$ in $B$ decays, $B \rightarrow K^{(*)} h'$ as discussed in~\cite{Filimonova:2019tuy}. Assuming visible decays with branching ratios as expected from Higgs mixing, \belletwo can reach a sensitivity down to a mixing angle of $\theta  \sim 10^{-5}$, assuming a final 
integrated luminosity of 50 ab$^{-1}$. 

\begin{figure}[tb]
  \centering
  \includegraphics[width=0.45\textwidth]{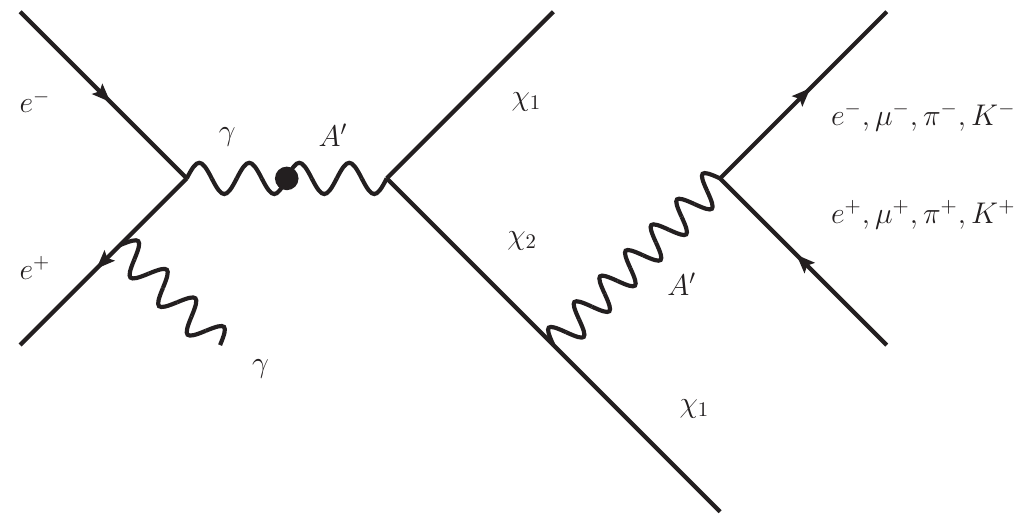}
  \includegraphics[width=0.45\textwidth]{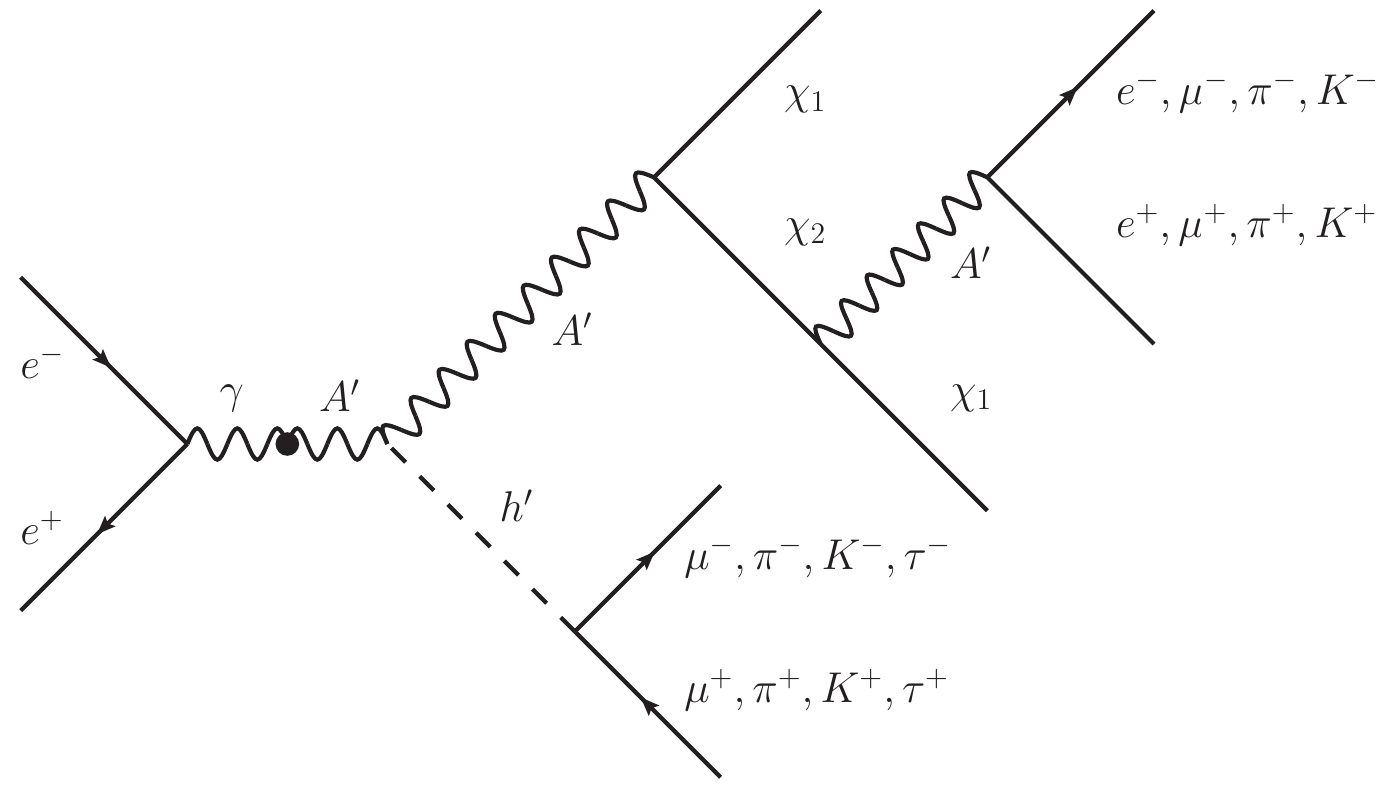}
  \caption{
    Feynman diagrams depicting the leading search channels for inelastic DM: $A'$ production in association with a single photon (left) and $A'$ production in association with a dark Higgs $h'$ (right) with subsequent decays into
    both visible and dark sector states.
    \label{fig:feynmandiagram}
  }
\end{figure}

Another possibility is direct production of the dark photon $A'$ through the kinetic mixing with the SM photon with subsequent decay into dark matter states $\chi_1$ and $\chi_2$ as depicted in 
Fig.~\ref{fig:feynmandiagram}. The production of $A'$ in association with a photon (left panel) has been discussed in detail in~\cite{Duerr:2019dmv}.
Depending on the decay length of $\chi_2$ the signature is either {\it (i)} a single photon with a displaced pair of charged particles and missing energy or  
{\it (ii)} a single photon with missing energy. Below we will implement these searches as described in~\cite{Duerr:2019dmv}.\footnote{In the current work we improve the description of the total 
$\chi_2$ decay width as described in the appendix.}

The process we will mainly concentrate on in this work includes a dark Higgs $h'$ in the intermediate state as depicted in the right panel of Fig.~\ref{fig:feynmandiagram}, leading to a signature
with missing energy and two pairs of charged particles. Specifically we will consider $\chi_2 \to \chi_1 \phi^+ \phi^-$ with $\phi=e,\mu,\pi, K$ and $h' \to \phi^+ \phi^-$ with $\phi=\mu,\pi, K, \tau$. 
The decay $h' \to e^+ e^-$ is very suppressed due to the small Yukawa coupling and charged hadrons other than $\pi,K$ are typically too short-lived to contribute to the signature. Pions and kaons behave similar to muons in the detector, so we will treat all of these particles identically in our analysis. To reduce backgrounds we will concentrate on the case where at least one pair of charged particles has a significant displacement. Before we enter a detailed discussion of the signature however, let us first describe the relevant aspects of the \belletwo experiment.

\subsection{The \belletwo experiment}
\begin{figure}[tb]
 \centering
 \includegraphics[angle=90,height=8cm]{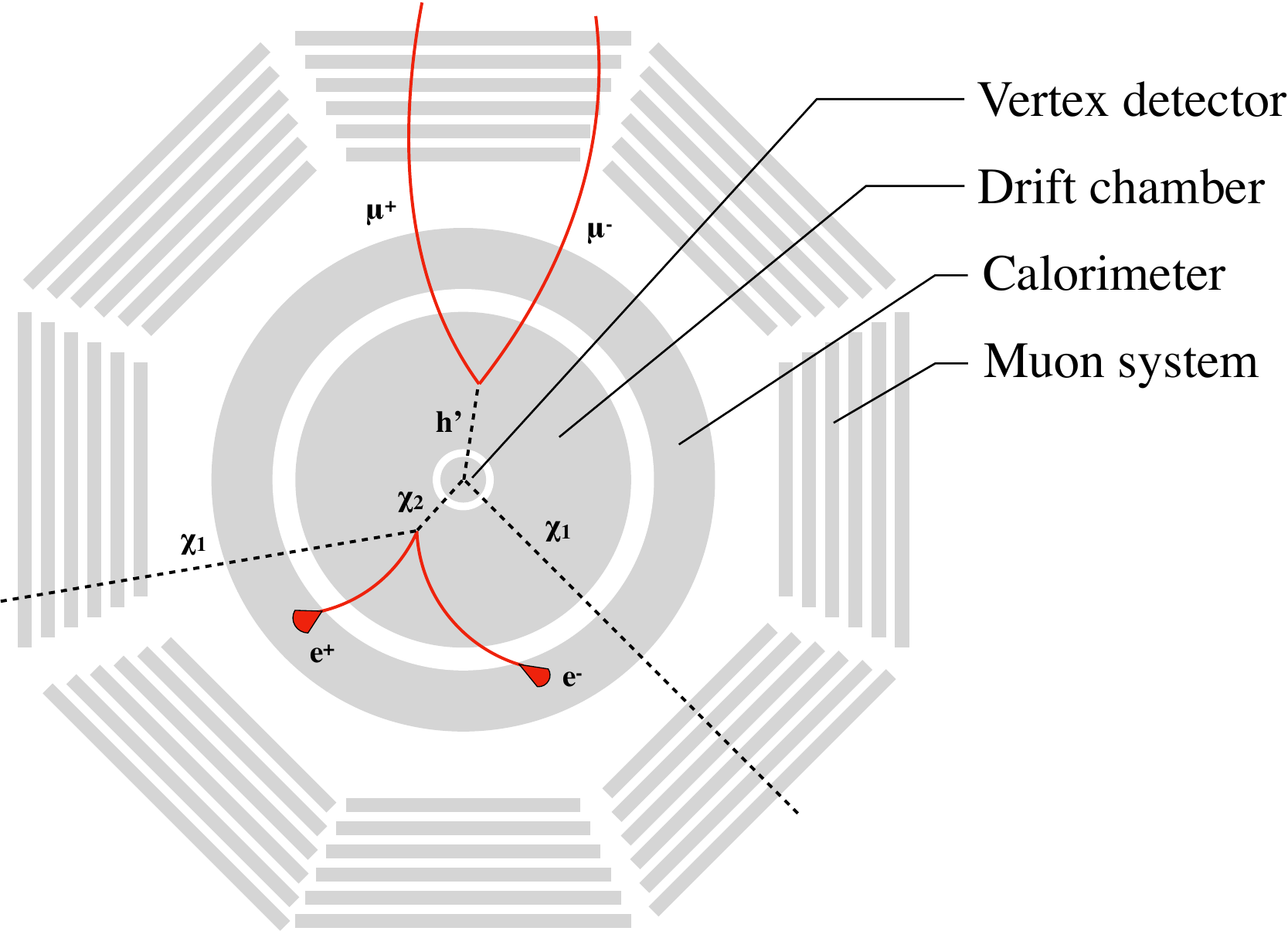}
 \caption{Schematic view of the \belletwo detector ($xy$-plane) and example displaced signature.}\label{fig:schematicdarkhiggs}
\end{figure}

The \belletwo experiment at the SuperKEKB accelerator is a next generation \bfactory~\cite{Abe:2010gxa} that started physics data taking in 2019. 
\superkekb is a circular asymmetric $e^+e^-$ collider with a nominal collision energy of $\sqrt{s} = \unit[10.58]{GeV}$ and a design instantaneous luminosity of \hbox{$8\times10^{35}$\,cm$^{-2}$ s$^{-1}$}. 

The \belletwo detector is a large-solid-angle magnetic spectrometer. 
 Particularly relevant for the searches described in this paper are the following sub-detectors: 
a tracking system that consists of six layers of vertex detectors (VXD), including two inner layers of silicon pixel detectors (PXD) and four outer layers of silicon vertex detectors (SVD), and a 56-layer central drift chamber (CDC) which covers a polar angle region of $(17-150)^{\circ}$. 
The electromagnetic calorimeter (ECL) comprising CsI(Tl) crystals with an upgraded waveform sampling readout for beam background suppression covers a polar angle region of $(12-155)^{\circ}$ and is located inside a superconducting solenoid coil that provides a 1.5\,T magnetic field. 
The ECL has inefficient gaps between the endcaps and the barrel for polar angles between $(31.3-32.2)^{\circ}$ and $(128.7-130.7)^{\circ}$. 
An iron flux-return is located outside of the magnet coil and is instrumented with resistive plate chambers and plastic scintillators to mainly detect $K^0_L$ mesons, neutrons, and muons (KLM) that covers a polar angle region of $(25-145)^{\circ}$. 

We study the \belletwo sensitivity for a dataset corresponding to an integrated luminosity of $\unit[100]{fb^{-1}}$ and $\unit[50]{ab^{-1}}$.
This dataset is expected to be recorded by \belletwo in early 2021 and by the end of \belletwo running around 2030, respectively.

 \subsection{Event Generation}

We implemented the model as specified in section~\ref{SEC:theory} into \texttt{FeynRules v2.3.32}~\cite{Alloul:2013bka} and generated a \texttt{UFO} model file~\cite{Degrande:2011ua}. To generate events for the process $e^+ e^- \to \chi_1 \chi_2 h'$ with subsequent decays of $h'$ and $\chi_2$ we employ \texttt{MadGraph5\textunderscore{}aMC@NLO v2.7.2}~\cite{Alwall:2014hca}. Specifically we simulate the decays 
$h' \to \mu^+ \mu^-$ and $\chi_2 \to \chi_1 l^+ l^-$ with $l=e,\mu$ so that 
the cross section that we obtain from \texttt{MadGraph} corresponds to
\begin{align}
    \sigma(e^+ e^- \to \chi_1 \chi_1 l^+ l^- \mu^+ \mu^-) = \sigma (e^+ e^- \to \chi_1 \chi_2 h') \times \text{BR}_{\chi_2 \to \chi_1 l^+ l^-} \times \text{BR}_{h' \to \mu^+ \mu^-}\,.
    \label{eq:cross_section}
\end{align}
To maximise the sensitivity of \belletwo we will however also be interested in other final states induced by the decays $\chi_2\to\chi_1\phi^+\phi^-$ with $\phi=e,\mu,\pi,K$ and $h' \to \phi^+\phi^-$ with $\phi=\mu,\pi,K,\tau$.
Given that pions and kaons behave similarly to muons as far as \belletwo is concerned, we do not simulate these particles in the final state explicitly but rather rescale the events with muons according to the relevant branching ratios. To this end we make use of the experimentally measured ratio $R(s)$ as described below. 
Similarly we rescale the events for decays into $\tau$ leptons.

As we concentrate on the region in parameter space in which $m_{A^\prime} > m_{\chi_1} + m_{\chi_2}$, the $A^\prime$ will never be on-shell in the $\chi_2$ decay and only three-body decays are possible. 
Nevertheless the $\chi_2$ branching fractions are largely determined by the $A'$ branching ratios (which we take from~\cite{Ilten:2018crw}).
To obtain the decay width of $\chi_2$, we numerically evaluate Eq.~\eqref{eq:chi2decayrate}
with and without the $R(s)$ factor to obtain the hadronic and leptonic contributions, respectively. We then feed the result of this calculation to \texttt{MadGraph}. 
While the total decay width of $\chi_2$ determines the decay length, we only take the partial decay widths which contribute to the desired final state into account when calculating the signal events. 
We conservatively assume that there is no contribution from pions and kaons above $\Delta = \unit[1.2]{GeV}$~\cite{Liu:2014cma}. 
The partial decay widths of $h'$, including hadronic final states, is taken from Ref.~\cite{Winkler:2018qyg}. To project out the charged final states we multiply the partial decay widths into 2 pions (kaons) by a factor 2/3 (2/4). Above 2 GeV multi-particle final states become relevant and we conservatively only consider the partial decay width into muons.

As in \cite{Duerr:2019dmv} we generate the events in the centre-of-mass frame with $\sqrt{s} = \unit[10.58]{GeV}$ and then boost and rotate them to the \belletwo laboratory frame.

\subsection{Signal selection}

We select events based on the radial vertex positions of the $h'$ and $\chi_2$ decay products (\textit{region selection}), the final state kinematics (\textit{kinematic selection}), and their trigger signatures (\textit{trigger selection}).
We consider different regions for the radial vertex positions as given in Table~\ref{TAB:regions_mumu} for $h'\to\phi^+\phi^-$ with $\phi=\mu,\pi,K$ and in Table~\ref{TAB:regions_tautau} for $h'\to\tau^+\tau^-$.
Note that $h'\to e^+e^-$ is suppressed and not considered further.
The different \textit{region selections} are defined based on the following arguments: Decays with $R<\unit[0.2]{cm}$ are very close to the nominal interaction point and will suffer from high SM backgrounds. 
The region $\unit[0.2]{cm} < R < \unit[0.9]{cm}$ is within the vacuum of the beam-pipe but sufficiently separated from the interaction point with no conversion backgrounds expected. 
The region $\unit[0.9]{cm} < R < \unit[17]{cm}$ includes the beam-pipe, support structures, the VXD, and the inner wall of the CDC with potentially large and complicated conversion and hadronic interaction backgrounds. 
We expect that those backgrounds can be removed only for non-electron final states, and we exclude $\chi_2\to \chi_1 e^+e^-$ decays in this region. 
$\unit[17]{cm} < R < \unit[60]{cm}$ covers the region inside the CDC with sufficiently high tracking efficiency and not much passive material.
For $\unit[60]{cm} < R < \unit[150]{cm}$ there will be enough activity in the detector (outer CDC, ECL, and inner KLM) to veto such final states in searches for invisible final states, but not enough information to reconstruct displaced vertices. 
$R>\unit[150]{cm}$ is only covered by the KLM with low efficiency for low momentum particles.

In addition, the events need to fulfil the \textit{kinematic selection} from Table~\ref{tab:selections}.
The angular selection is also applied to prompt decays. Finally an event must pass at least one of the \textit{trigger selection} that are explained in Sec.~\ref{SEC:triggers}.
While these selections are motivated by the performance shown in \cite{Kou:2018nap}, we note that a full study of all possible backgrounds is beyond the scope of this work. For all sensitivity predictions, we assume zero background after selections.

\begin{table*}[htb]
  \centering
\caption{\label{TAB:regions_mumu} Decay vertex regions for different values of radial displacement for the case $h'\to\phi^+\phi^-$ with $\phi=\mu, \pi, K$ used for the \textit{region selection}: In the gray region all $\chi_2$ final states are considered, i.e.\ $\chi_2\to\chi_1\phi^+\phi^-$ with $\phi=e,\mu,\pi,K$ while in the light blue region only $\chi_2\to \phi^+\phi^-$ with $\phi=\mu,\pi,K$ are considered due to significant pair conversion backgrounds for electrons.}
\begin{tabular}{c|ccccc} 
 \diagbox[innerwidth=12em]{$h'\to\phi^+\phi^-$}{$\chi_2\to \chi_1 \phi^+\phi^-$} & $<0.2$\,cm & 0.2--0.9\,cm & 0.9--17\,cm & 17--60\,cm &  $>\,$60\,cm\\
\hline
$<0.2$\,cm  &  & \cellcolor[HTML]{C0C0C0}\textbf{} & \cellcolor[HTML]{82CAFA}\textbf{} & \cellcolor[HTML]{C0C0C0}\textbf{} &  \\
0.2--0.9\,cm  & \cellcolor[HTML]{C0C0C0}\textbf{} & \cellcolor[HTML]{C0C0C0}\textbf{} & \cellcolor[HTML]{82CAFA}\textbf{} & \cellcolor[HTML]{C0C0C0}\textbf{} &  \\
0.9--17\,cm  & \cellcolor[HTML]{C0C0C0}\textbf{} & \cellcolor[HTML]{C0C0C0}\textbf{} & \cellcolor[HTML]{82CAFA}\textbf{} & \cellcolor[HTML]{C0C0C0}\textbf{} &  \\
17--60\,cm  & \cellcolor[HTML]{C0C0C0}\textbf{} & \cellcolor[HTML]{C0C0C0}\textbf{} & \cellcolor[HTML]{82CAFA}\textbf{} & \cellcolor[HTML]{C0C0C0}\textbf{} & \\
$>\,60$\,cm  &  &  &  &  & \\
\hline
\end{tabular}
\end{table*}

\begin{table*}[htb]
  \centering
\caption{\label{TAB:regions_tautau} Decay vertex regions for different values of radial displacement for $h'\to\tau^+\tau^-$ used for the \textit{region selection}: 
In the gray region all $\chi_2$ final states are considered, i.e.\ $\chi_2\to\phi^+\phi^-$ with $\phi=e,\mu,\pi,K$ while in the light blue region only $\chi_2\to \phi^+\phi^-$ with $\phi=\mu,\pi,K$ are considered.
}
\begin{tabular}{c|ccccc} 
  \diagbox[innerwidth=12em]{$h'\to\tau^+\tau^-$}{$\chi_2\to \chi_1 \phi^+\phi^-$} & $<0.2$\,cm & 0.2--0.9\,cm & 0.9--17\,cm & 17--60\,cm & $> 60$\,cm\\
 \hline
$<0.2$\,cm  &  &  &  &  &  \\
0.2--0.9\,cm  &  &  &  &  &  \\
0.9--17\,cm  & \cellcolor[HTML]{C0C0C0}\textbf{} & \cellcolor[HTML]{C0C0C0}\textbf{} & \cellcolor[HTML]{82CAFA}\textbf{} & \cellcolor[HTML]{C0C0C0}\textbf{} &  \\
17--60\,cm  & \cellcolor[HTML]{C0C0C0}\textbf{} & \cellcolor[HTML]{C0C0C0}\textbf{} & \cellcolor[HTML]{82CAFA}\textbf{} & \cellcolor[HTML]{C0C0C0}\textbf{} &  \\
$> 60$\,\,cm  &  &  &  &  &  \\
 \hline
\end{tabular}
\end{table*}

For $h'\to\tau^+\tau^-$ events we modify our \textit{region selection} as follows:
We assume that all $\tau$ decay modes are usable for the analysis.
We require more displacement compared to $h'\to\mu^+\mu^-/h^+h^-$ because the $\tau^+\tau^-$ vertex is harder to resolve experimentally.
For low momentum $\tau$ decays, the daughter tracks may not point back to the interaction point which will require improved pattern recognition for the analysis.
Note that for the values of $\Delta$ used in this work, the decays of $\chi_2\to\chi_1\tau^+\tau^-$ are kinematically not possible.

\begin{table}[bt]
\centering
 \caption{\textit{Kinematic selections} used in our analysis. 
 \label{tab:selections}}
 \begin{tabular}{ll}
  cut on  & value \\
  \hline \hline
   \multirow{2}{8em}{decay vertex} & (i) $\unit[-55]{cm} \leq z \leq \unit[140]{cm}$\\ 
                                         & (ii) $ 17^\circ\leq \theta_\text{lab} \leq 150^\circ$ \\
   \hline
   \multirow{3}{8em}{electrons} &  (i) both $p(e^+)$ and $p(e^-) > \unit[0.1]{GeV}$\\
             &  (ii) opening angle of pair $> 0.1$ \,rad\\
             &  (iii) invariant mass of pair $m_{ee}> \unit[0.03]{GeV}$\\
             \hline
   \multirow{4}{8em}{$\mu, \pi, K, \tau$}     &  (i) both $p_\text{T}(\mu^+)$ and $p_\text{T}(\mu^-) > \unit[0.05]{GeV}$\\ 
             &  (ii) opening angle of pair $> 0.1$ \,rad\\
             &  (iii) invariant mass of pair $m_{ll} > \unit[0.03]{GeV}$\\
             &  (iv)  $m_{ll} < \unit[0.480]{GeV}$ or $m_{ll} > \unit[0.520]{GeV}$ \\
   \hline \hline
 \end{tabular}
\end{table}

\subsection{Triggers}
\label{SEC:triggers}

We consider the following triggers, to be able to cover the various interesting regions discussed below. 
The triggers are similar to those described~\cite{Duerr:2019dmv}, but the criteria have been refined to better match the trigger algorithms in \belletwo.
The trigger conditions are approximately the combined \belletwo hardware and software triggers.
\begin{itemize}
\item \textbf{2 GeV energy:} Requires at least one calorimeter cluster with $E_\text{CMS} > \unit[2]{GeV}$ and {$22^\circ < \theta_\text{lab} < 139.3^\circ$}. Only the electrons coming from the $\chi_2$ decay can potentially deposit this amount of energy (even they hardly ever will pass this trigger), whereas $\mu, \pi, K$ are not expected to trigger at all.
This trigger does not work for a displacement larger than the radius of the electromagnetic calorimeter, which we take to be $R_\text{ECL} = \unit[1.35]{m}$. 

\item \textbf{Three isolated clusters:} Requires at least three isolated calorimeter clusters with a minimum distance of $d_\text{min} = \unit[30]{cm}$.  At least one of the three clusters needs to have {$E_\text{lab} > \unit[0.5]{GeV}$} (which can only be deposited by the electrons from the $\chi_2$ decay), and there need to be two additional clusters with $E_\text{lab} > \unit[0.18]{GeV}$, which can be either electrons or $\mu, \pi, K$. All three clusters need to have {$18.5^\circ < \theta_\text{lab} < 139.3^\circ$}. This trigger will potentially be prescaled (i.e.\ only a fraction of these events will actually be kept) for the full Belle II data set. 

\item \textbf{Four isolated clusters:} Requires at least four isolated calorimeter clusters with a minimum distance of $d_\text{min} = \unit[30]{cm}$ with {$E_\text{lab} > \unit[0.18]{GeV}$}, which can be either electrons or $\mu, \pi, K$. All four clusters need to have {$18.5^\circ < \theta_\text{lab} < 128.7^\circ$}. At least one of the four clusters needs to have $E_\text{lab} > \unit[0.3]{GeV}$.

\item \textbf{Two tracks:} Requires two tracks with a transverse momentum $p_T > \unit[300]{MeV}$ each and {$38^\circ < \theta_\text{lab} < 127^\circ$}, as well as an azimuthal opening angle at the interaction point in the lab system $ \Delta \varphi > 90^\circ$. We assume this trigger is not efficient beyond a radius of $R_\text{max} = \unit[17]{cm}$. 

\item \textbf{Three tracks:} Same conditions as the \textbf{two tracks} triggers, but without requirement on $ \Delta \varphi$.

\item \textbf{1 GeV $\boldsymbol{E}$ sum:} Requires that the sum of all clusters with $E_\text{lab} > \unit[100]{MeV}$ and $27^\circ < \theta_\text{lab} < 128^\circ$ (covers the barrel and outer forward endcap) is larger than 1 GeV. $\mu, \pi, K$ contribute \unit[200]{MeV} to this sum if their momentum is sufficiently large.

\item \textbf{Displaced vertex:} Requires at least one displaced vertex in the event with $\unit[0.9]{cm} < R_{xy} < \unit[60]{cm}$, and a transverse momentum of the corresponding particles of $p_T > \unit[100]{MeV}$ each.  While this trigger is currently not implemented in \belletwo, we expect that no hardware modifications are needed, and that dedicated algorithms can be implemented in the firmware of the existing trigger. 

\end{itemize}

For $h'\to\tau^+\tau^-$ events we simplify our \textit{region selection} as follows:
We assume 100\,\% trigger efficiency for $h'\to\tau^+\tau^-$ since about 70\,\% of all $\tau$ decay modes include at least one electron, one $\pi^0$, or three charged hadrons.
For $\tau$ pairs this results in about 90\,\% final states with at least one such final state where trigger efficiency will generally be high. 
A detailed study is beyond the scope of this work given the multitude of possible final states.

\section{Results}
\label{sec:results}

Let us now come to a discussion of the expected sensitivities at \belletwo with respect to the different possible signatures. As the model exhibits seven free parameters, a full evaluation of the different sensitivities in the entire parameter space would require a global scan and is beyond the scope of this work. Instead we show some exemplary parameter planes which illustrate the typical strength of different searches. 

Comparing the two different Feynman diagrams in Fig.~\ref{fig:feynmandiagram} we observe that they have a somewhat different dependence on the model parameters, with the cross section for the case with an associated photon scaling as $\sigma \propto \epsilon^2 \alpha^2$ while the signature with the dark Higgs $h'$ in the final state scaling as $\sigma \propto \epsilon^2 \alpha \alpha_D$.
For both cases the total production cross section is dominated by on-shell production of $A'$,
with subsequent decays $A' \to \chi_1 \chi_2$ and $\chi_2 \to \chi_1 \phi^+ \phi^-$ with $\phi$ some SM state.

\begin{figure*}[t]
\centering
\includegraphics[width=0.39\textwidth]{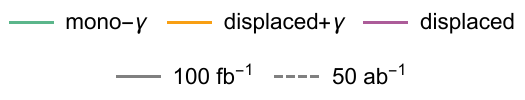}\\
\includegraphics[width=0.49\textwidth]{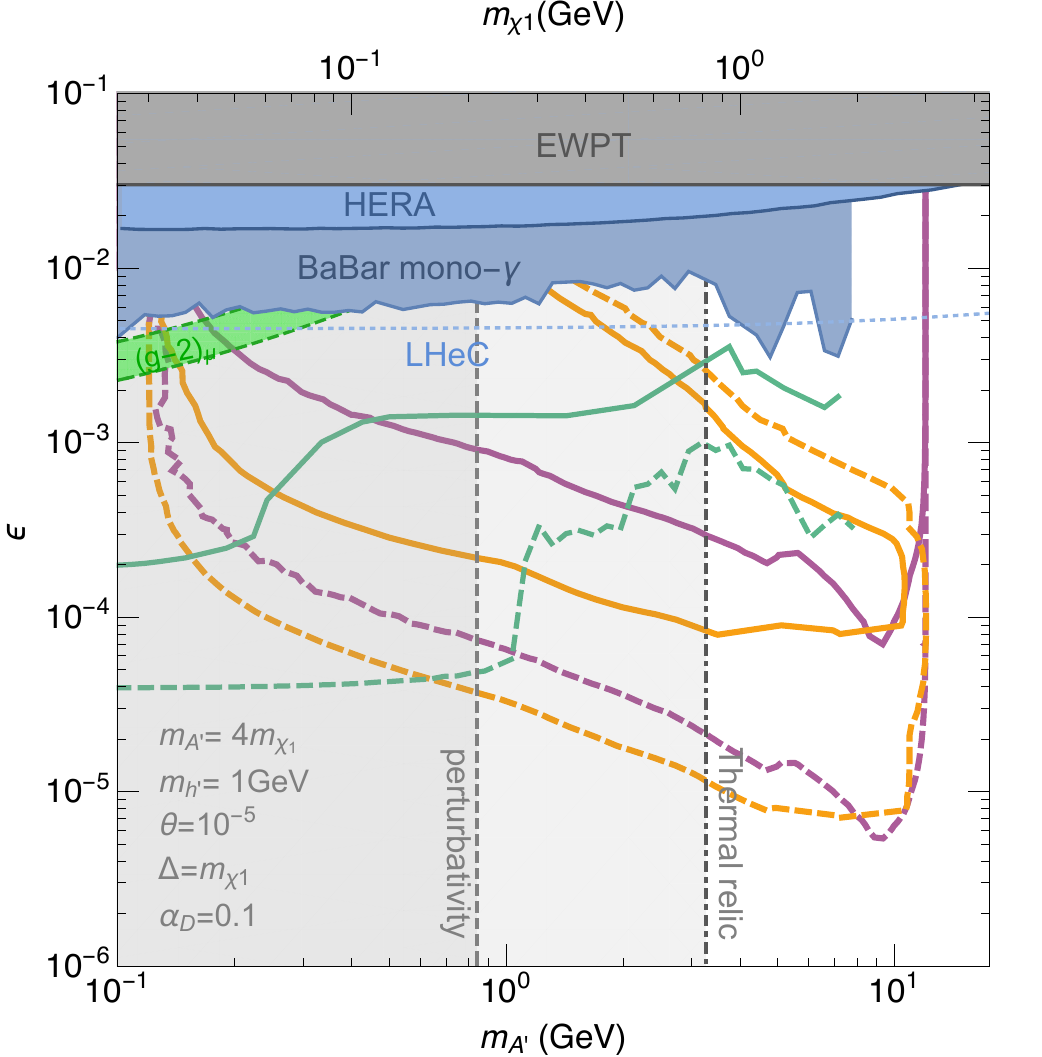}
\includegraphics[width=0.49\textwidth]{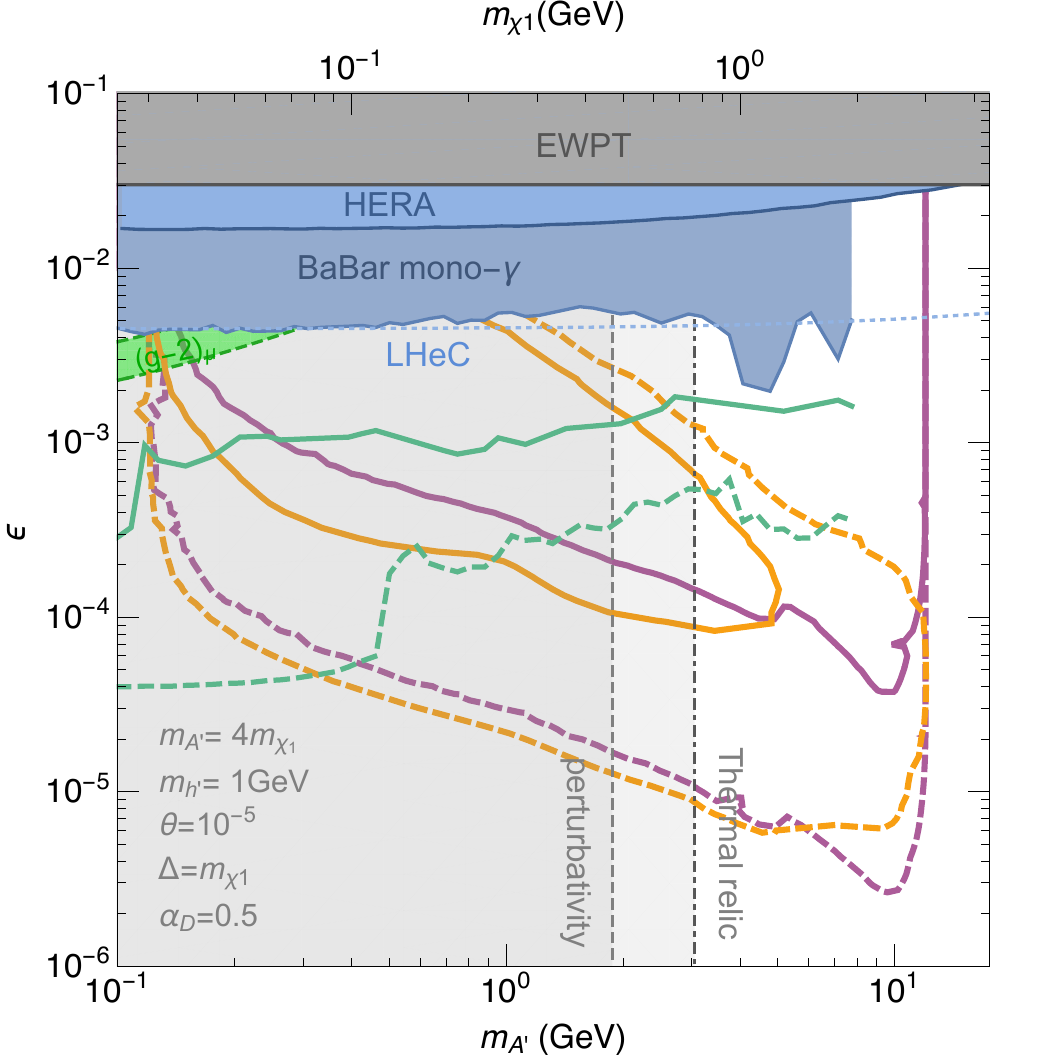}
\caption{Expected sensitivities of the different searches at \belletwo in the $\epsilon - m_{A'}$ parameter plane for integrated luminosities of 100 fb$^{-1}$ (solid lines) and 50 ab$^{-1}$ (dashed lines). Left plot is for $\alpha_D = 0.1$, right plot for $\alpha_D = 0.5$.  
}
\label{fig:epsilon}
\end{figure*}

In Fig.~\ref{fig:epsilon} we show the sensitivities of the different searches at \belletwo in the $\epsilon - m_{A'}$ parameter plane for integrated luminosities of 100 fb$^{-1}$ (solid lines) and 50 ab$^{-1}$ (dashed lines). The other parameters are fixed as indicated in the figures. We show 90\% C.L. limits for all signatures analysed in this work, i.e.\ for the monophoton as well as the two displaced signatures at  \belletwo.
Existing bounds come from electroweak precision tests (EWPT)~\cite{Hook:2010tw} and from HERA measurements~\cite{Kribs:2020vyk} as well as from the \babar monophoton search~\cite{Abe:2010gxa}. As described in \cite{Duerr:2019dmv} we run Monte Carlo scans to take into account the fact that only a fraction of the events will pass the monophoton selection criteria, resulting in a significantly weaker bound from \babar for the given parameters. For the rather large value of $\Delta$ and $\epsilon$ almost all $\chi_2$ particles will decay within the detector and the remaining limit from the monophoton signature is due to the non-zero probability that the particles produced in the $\chi_2$ decay travel in the direction of the beam pipe such that they will not be reconstructed.

The sensitivity of \belletwo towards the monophoton signature (green) is significantly improved compared to BaBar due to a more hermetic calorimeter.
To obtain the monophoton sensitivity for $\unit[100]{fb^{-1}}$ and $\unit[50]{ab^{-1}}$ we rescale the published sensitivity for $\unit[20]{fb^{-1}}$ using that the expected sensitivity $S(\epsilon) \propto \sqrt[4]{\mathcal{L}}$.\footnote{The assumptions under which such a rescaling is valid are discussed in detail in~\cite{Duerr:2019dmv}.} We then perform a second rescaling as above using Monte Carlo runs to account for $\chi_2$ decays and corresponding acceptances within the detector.
We observe that for small values of $m_{A'}$ the sensitivity is as good as for the usual monophoton search as basically all $\chi_2$ particles decay outside the detector. For larger $m_{A'}$ this is no longer true and we observe a significant weakening (which is delayed for larger luminosities due to the smaller values of $\epsilon$ and therefore larger $\chi_2$ decay lengths).

In orange we show the sensitivity due to the signature with a single photon and a displaced pair of charged particles (denoted by `displaced+$\gamma$' in the figure legend). We observe that there is very good sensitivity towards large dark photon masses $m_{A'}$ and rather small values of $\epsilon$.
In violet we show the corresponding sensitivity for the signature with two pairs of charged particles, where we require at least one of those to have a
non-zero displacement (denoted by `displaced' in the figure legend).
While the typical sensitivity is very similar to the `displaced+$\gamma$'
signature, it extends to large values of $\epsilon$ which are not covered by any other signature. The reason is that we can allow for prompt $\chi_2$ decay in this case as the decay products of the dark Higgs $h'$ are basically always displaced. We further note that the constraints extend significantly into the off-shell regime with dark photon masses $m_{A'} \lesssim \unit[12]{GeV}$ for $m_{h'}=\unit[1]{GeV}$.

Because the relic density depends primarily on the process $\chi_1 \chi_1 \to h^\prime h^\prime$, the thermal relic target does not depend on $\varepsilon$ or $\theta$ in this case.
For comparison we also show the preferred parameter region in which the predicted anomalous magnetic moment of the muon~\cite{Pospelov:2008zw,Mohlabeng:2019vrz,Abi:2021gix} is within the $2 \sigma$ range of its experimentally measured value.\footnote{For experimentally allowed values of the Higgs mixing parameter $\theta$ the impact of the dark Higgs on the predicted value of $g-2$ is negligible, see e.g.~\cite{Chen:2015vqy}.}
We observe that while for the dark Higgs masses $m_{h'}$ we are interested in here the region corresponds to a non-perturbative quartic dark Higgs coupling, this model offers a viable explanation for
correspondingly smaller values of $m_{h'}$.

\begin{figure*}[t]
\centering
\includegraphics[width=0.39\textwidth]{./figs/fig9.pdf}\\
\includegraphics[width=0.49\textwidth]{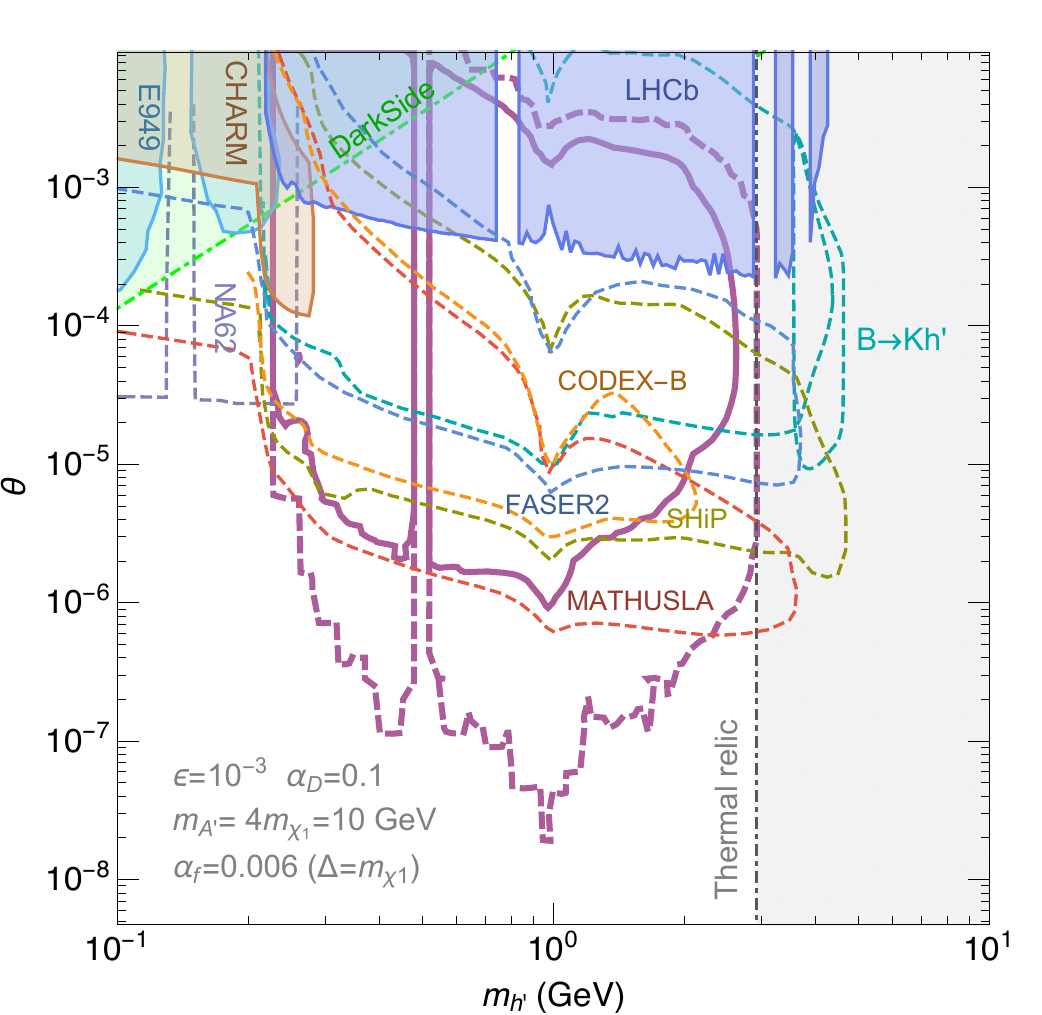}
\includegraphics[width=0.49\textwidth]{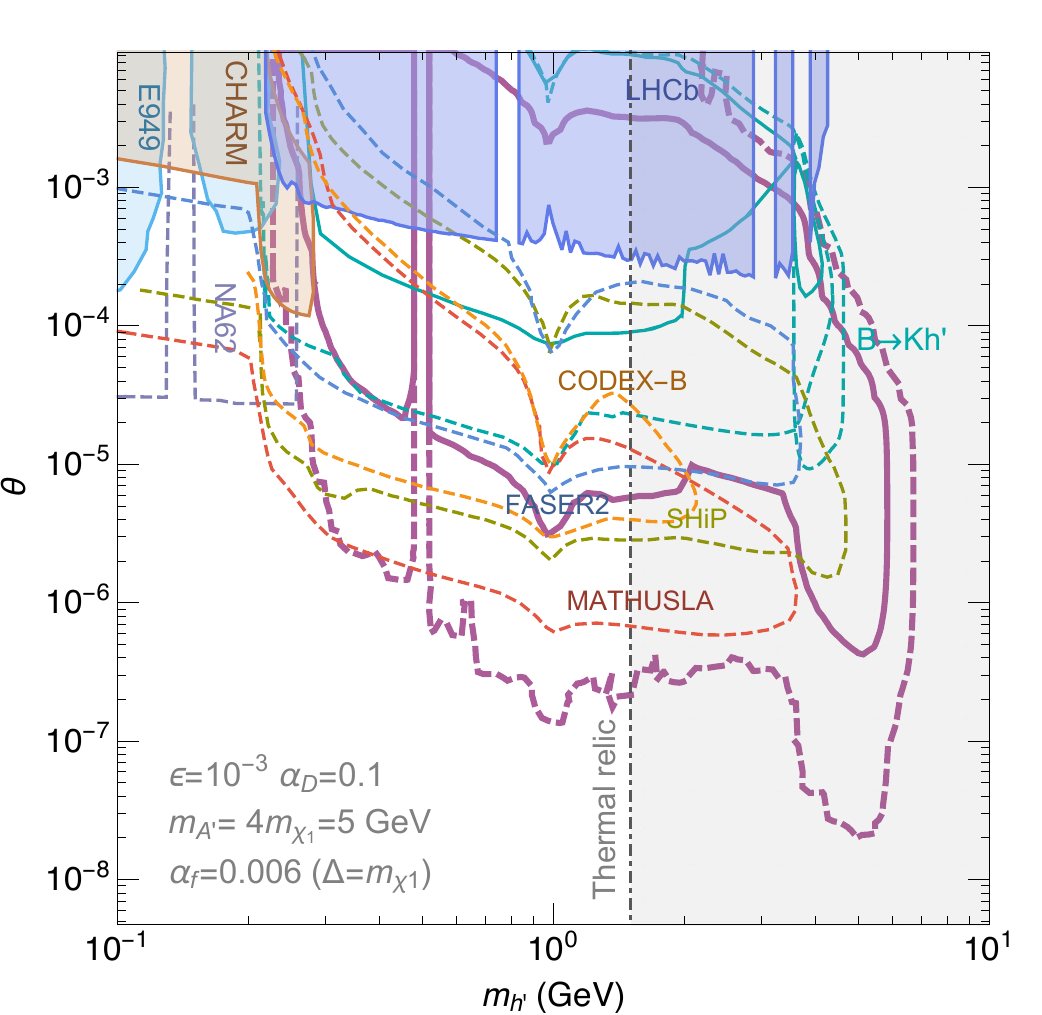}
\caption{Expected sensitivities of the different searches at \belletwo in the $\theta - m_{h'}$ parameter plane for integrated luminosities of 100 fb$^{-1}$ (solid lines) and 50 ab$^{-1}$ (dashed lines). We also show current limits from DarkSide~\cite{Agnes:2018ves}, LHCb, CHARM and E949 (taken from \cite{Winkler:2018qyg}) as well as a number of expected sensitivities of proposed future searches as shown in \cite{Winkler:2018qyg} and \cite{Ariga:2018uku}.
}
\label{fig:theta}
\end{figure*}

In Fig.~\ref{fig:theta} we show the limits in the $\theta - m_{h'}$ parameter plane. Here general searches for dark scalars mixing with the SM Higgs boson are relevant and we show results from LHCb, CHARM and E949 as given in~\cite{Winkler:2018qyg}.
We also show limits from direct dark matter searches, taking into account the fact that for the regions in parameter space where $\chi_1$ does not make up all the DM (to the left of the `thermal relic' line), the limits have to be rescaled with a factor $\Omega_{\chi_1} h^2/0.12$. 

Regarding future sensitivities we show estimates for NA62 (as given in~\cite{Bondarenko:2019vrb}), SHiP (as given in~\cite{Winkler:2018qyg})
and a possible \belletwo search for the rare decay $B\to K h'$~\cite{Filimonova:2019tuy}. For the given set of parameters the monophoton as well as the `displaced+$\gamma$' searches are not sensitive.
The signature associated with the dark Higgs however is sensitive down to very small values of the mixing angle $\theta$. This remarkable sensitivity can be understood from the fact that the production cross section is large and does not depend on $\theta$. The lower boundary of the sensitivity is 
therefore just given by the maximal $h'$ decay length which still allows
for 2.3 events to decay within the sensitive region of the detector.
The maximal decay length which \belletwo can be sensitive to corresponds to more than $10^5$m.

\begin{figure*}[t]
\centering
\includegraphics[
width=0.39\textwidth]{./figs/fig9.pdf}\\
\includegraphics[width=0.49\textwidth]{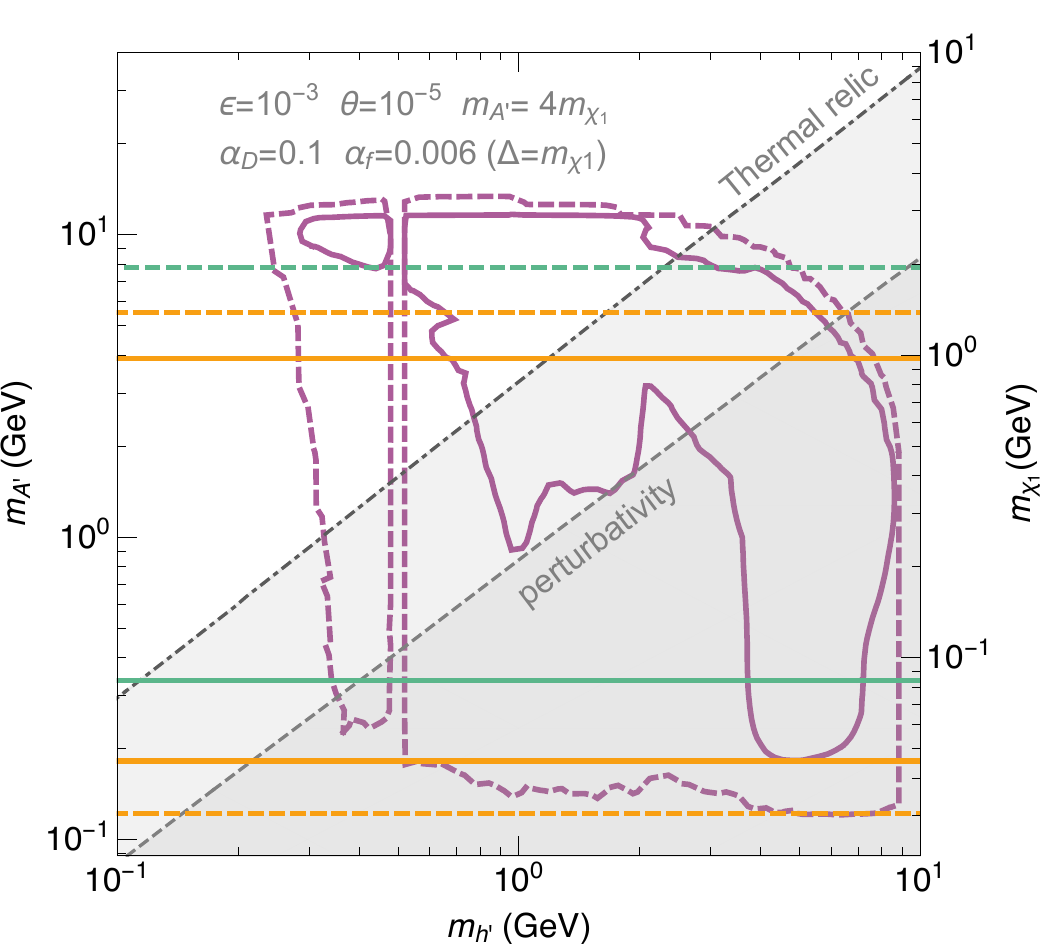}
\includegraphics[width=0.49\textwidth]{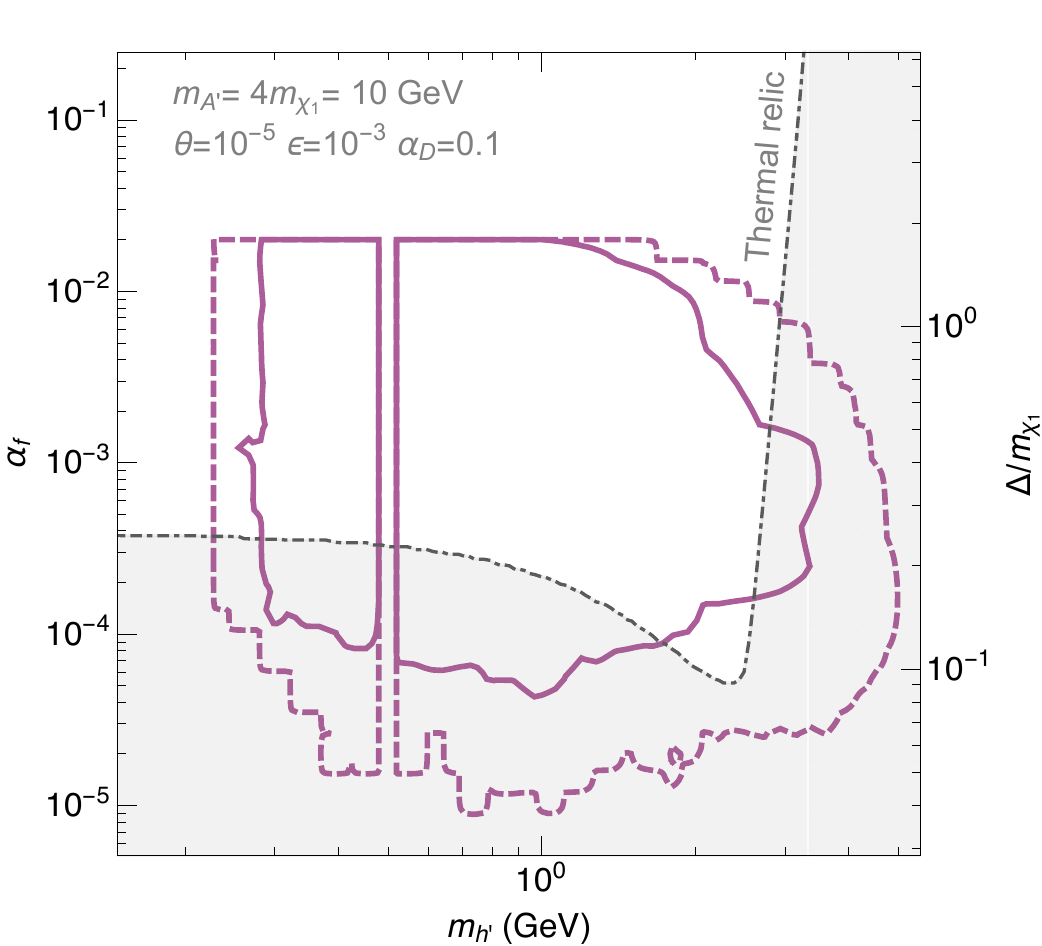}
\caption{Expected sensitivities of the different searches at \belletwo in the (left) $m_{h'}-m_{A'}$ plane and in the (right) $m_{h'}-\alpha_f$ plane for integrated luminosities of 100 fb$^{-1}$ (solid lines) and 50 ab$^{-1}$ (dashed lines). 
}
\label{fig:africaANDrelic}
\end{figure*}

\begin{figure*}[t]
\centering
\includegraphics[width=0.49\textwidth]{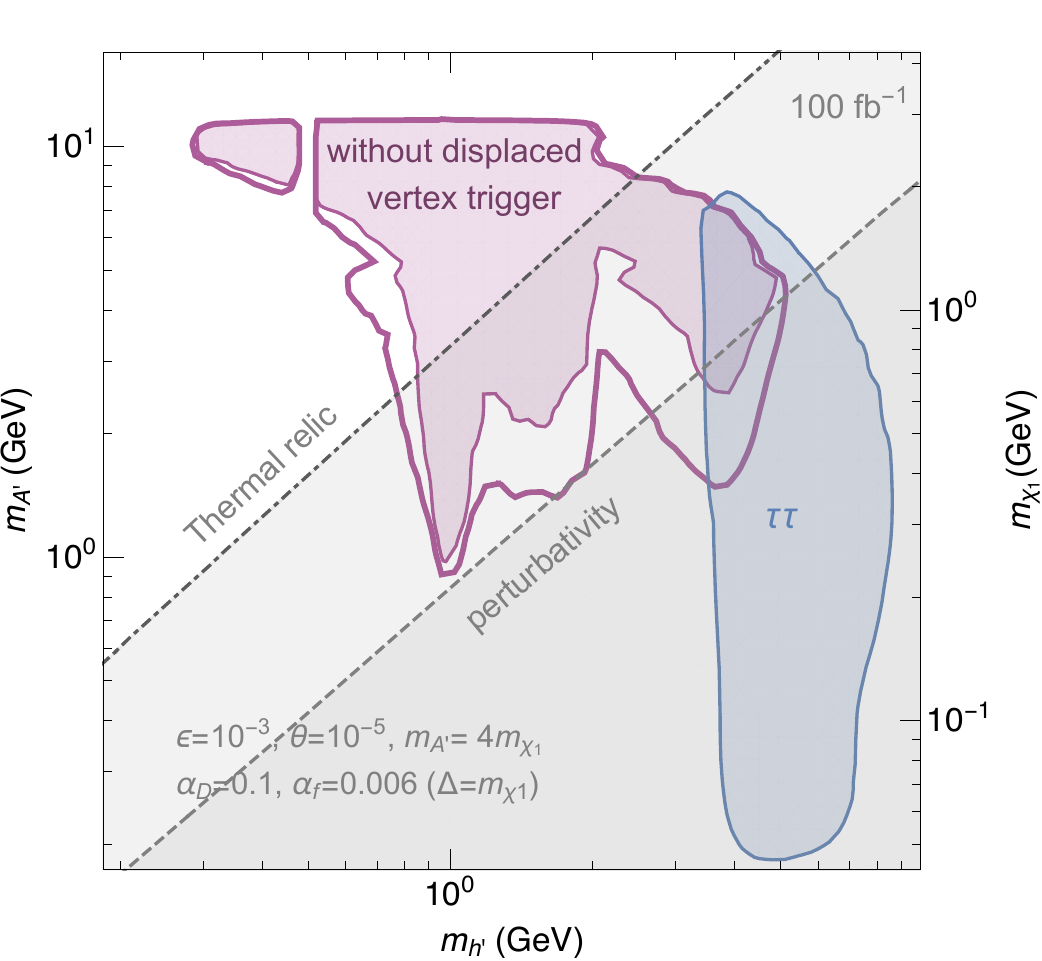}
\includegraphics[width=0.49\textwidth]{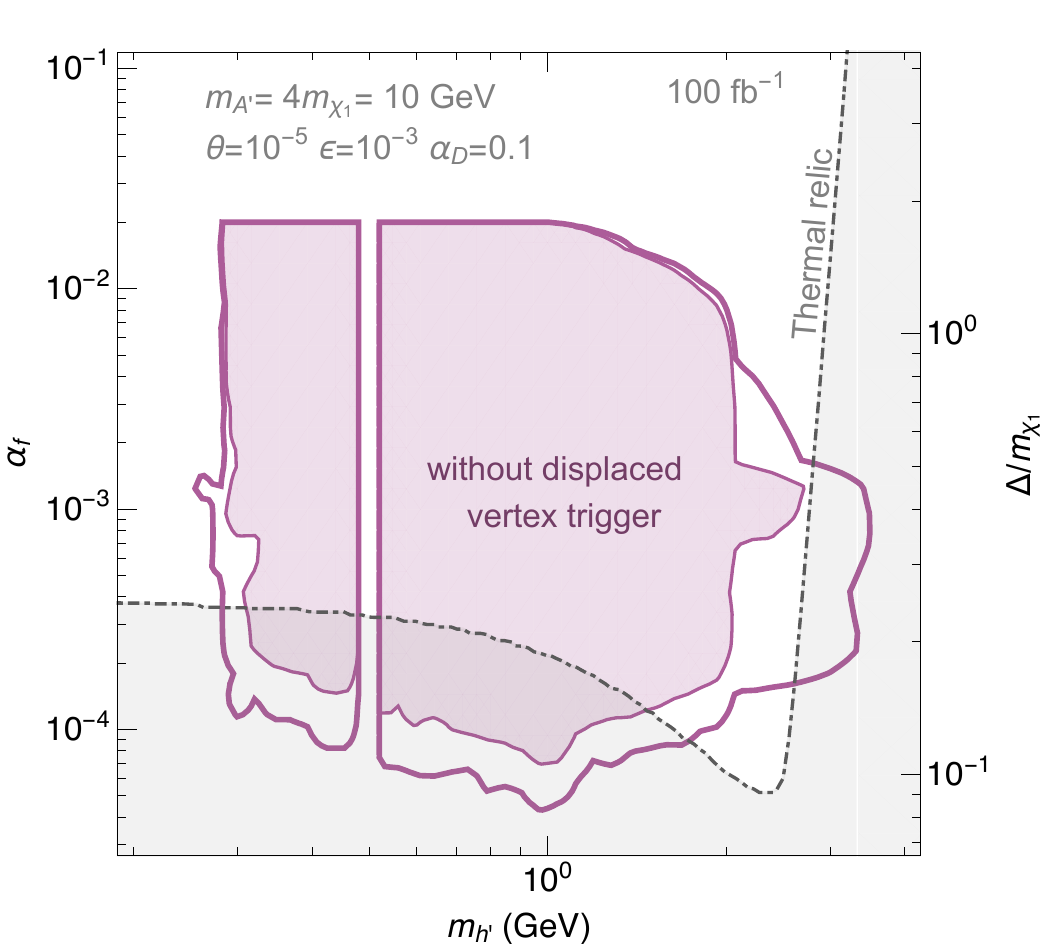}
\caption{Expected sensitivities of the displaced search at \belletwo in the $m_{h'}-m_{A'}$ plane (left) and in the $m_{h'}-\alpha_f$ plane (right) for integrated luminosity of 100 fb$^{-1}$. The filled regions correspond to the sensitivity without invoking a displaced vertex trigger. In addition we show the region in which the $\tau\tau$ region contributes to the overall sensitivity.
}
\label{fig:displaced_vertex}
\end{figure*}
In Fig.~\ref{fig:africaANDrelic} we show the sensitivities of the different \belletwo searches in the $m_{h'}-m_{A'}$ plane (left) and in the $m_{h'}-\alpha_f$ plane (right). Note that we assume that in the parameter region around $m_{h'}\sim \unit[0.5]{GeV}$ the search does not have any sensitivity due to large $K_S$ backgrounds (see the selection cuts in Tab~\ref{tab:selections}), explaining the gap in our sensitivity. 
In Fig.~\ref{fig:displaced_vertex} we show the same planes as in Fig.~\ref{fig:africaANDrelic} but restrict ourselves to the case of $\unit[100]{fb^{-1}}$ to show more details of how the sensitivity region
depends on the assumption of the presence of a displaced vertex trigger.
We see that a displaced vertex trigger could significantly extend the reach in some regions of parameter space while in others there is only a mild improvement. Experimentally, a displaced vertex track trigger would be orthogonal to the calorimeter triggers and will hence provide a way to measure the trigger efficiency.

\section{Conclusion}

In this work we studied possible signatures at \belletwo of a simple model for light thermal inelastic dark matter which is fully consistent with all cosmological probes as well as direct and indirect dark matter detection. 
We extend previous studies of inelastic dark matter by carefully 
analysing the effects of a dark Higgs boson $h'$, which is naturally present in the low energy particle spectrum to explain the mass splitting $\Delta$ between the DM state $\chi_1$ and its heavier twin $\chi_2$ as well as the mass of the dark photon $m_{A'}$.
One straightforward consequence of the presence of the dark Higgs $h'$ is that {\it elastic} scattering between $\chi_1$ and nuclei is possible even at tree-level (making the term {\it inelastic} DM something of a misnomer). Nevertheless, the resulting scattering cross section is still rather small due to the small couplings involved and typically not competitive with limits from colliders. 

A prominent signature at \belletwo which arises from dark Higgs particles  $h'$ produced in association with dark matter $\chi_1$ consists of two pairs of (displaced) charged particles together with missing momentum.
We find that the sensitivity of \belletwo to the underlying model parameters is highly complementary to that from monophoton searches, while an independent signature with a single photon, one pair of charged particles and missing momentum as studied in~\cite{Duerr:2019dmv} gives very similar sensitivity in large regions of parameter space. 
The signature involving a dark Higgs however provides sensitivity also to large values of $\epsilon$ which are not covered by any other signature.
Overall it appears not unlikely that both signatures may be discovered almost simultaneously at \belletwo, providing a unique signature correlation for this scenario.
We also point out that some regions of parameter space will not be covered with the current experimental configuration and that a displaced vertex trigger would be highly beneficial to increase the sensitivity to this scenario. 

\section{Note added}
\begin{figure*}[t]
\centering
\includegraphics[width=0.39\textwidth]{./figs/fig9.pdf}\\
\includegraphics[width=0.49\textwidth]{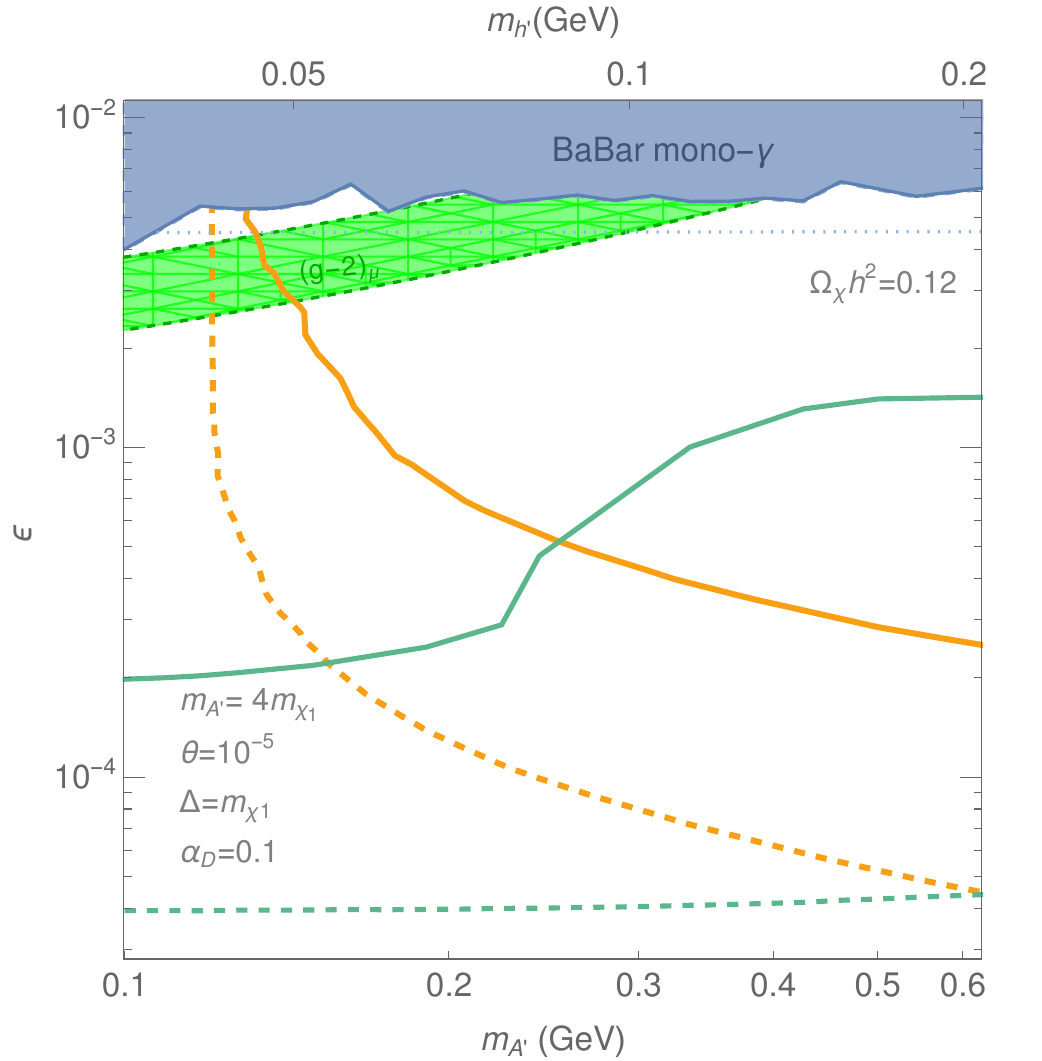}
\caption{Expected sensitivities of the different searches at \belletwo in the $\epsilon - m_{A'}$ parameter plane for integrated luminosities of 100 fb$^{-1}$ (solid lines) and 50 ab$^{-1}$ (dashed lines), similar to Fig.~\ref{fig:epsilon}, but zoomed into the region in which the anomalous magnetic moment of the muon can be explained (green). The dark Higgs mass $m_{h'}$ is chosen such that the correct relic abundance is achieved for all values of $m_{A'}$ (see upper axis for required value of $m_{h'}$). The preferred range for the anomalous magnetic moment in this figure is calculated including the recent observation from~\cite{Abi:2021gix}.
}
\label{fig:gm2}
\end{figure*}

After the completion of this work the E989 experiment at Fermilab presented a new result of the muon anomalous magnetic moment~\cite{Abi:2021gix} which is in agreement with the previous result at BNL~\cite{Bennett:2006fi} and has increased the tension with the SM prediction to 4.2$\sigma$.\footnote{A recent evaluation of the leading hadronic contribution to the muon magnetic moment from lattice QCD~\cite{Borsanyi:2020mff} suggests that this discrepancy may in fact be significantly smaller.}
In this context we would like to point out that while a \textit{kinetically mixed} dark photon which decays purely visibly (or invisibly) is excluded as a possible explanation of this finding,
in our scenario this discrepancy can readily be resolved.
In fact, for the range of dark Higgs masses $m_{h'}$ we consider here, the perturbativity constraint shown in Fig.~\ref{fig:epsilon} no longer applies while simultaneously the correct dark matter relic abundance can be achieved for all values of $m_{A'}$, cf.~Fig.~\ref{fig:gm2}.\footnote{For masses $m_{A'} \simeq 0.1$~GeV there may also be complementary constraints from fixed target experiments such as E137, cf.\ Ref~\cite{Izaguirre:2017bqb,Mohlabeng:2019vrz}.}
Even more interestingly, the corresponding region in parameter space will be fully tested by the \belletwo experiment in the near future, either excluding this possible resolution or discovering a clear signal
for physics beyond the SM.

\acknowledgments

We would like to thank Luc Darme, Felix Kahlh\"ofer and Jure Zupan for discussions, Martin Winkler for providing us with the exclusion lines presented in~\cite{Winkler:2018qyg}, Felix Kling for providing us with the exclusion lines presented in \cite{Ariga:2018uku} and Anastasiia Filimonova, Ruth Sch\"afer, and Susanne Westhoff for discussions and for providing us with the exclusion lines presented in~\cite{Filimonova:2019tuy}.
This work is funded by the Deutsche Forschungsgemeinschaft (DFG) through Germany's Excellence Strategy -- EXC 2121 ``Quantum Universe'' -- 390833306, by the ERC Starting Grant ``NewAve'' (638528), the Helmholtz (HGF) Young Investigators Grant No.\ VH-NG-1303, and the Science and Technology Facilities Council~(STFC) grant ST/P000770/1. C.G.C. is supported by the Alexander von Humboldt Foundation.

\appendix

\section{The decay of the excited DM state into hadronic channels}

\newcommand{\tm}{\textit{muons}}

\textbf{Cross sections.} If the annihilation of $\chi_1$ and $\chi_2$  into a final state ${\cal F}$ is induced by the s-channel exchange of a dark photon, its amplitude can be cast as
\begin{align}
{\cal M} \left(\chi_1 \chi_2 \to {\cal F}\right) = e g_X  \epsilon^2 \overline{v_2} \gamma^\alpha u_1 \left(\frac{-g_{\alpha\beta}+P_\alpha P_\beta/m_{A'}^2}{P^2-m_{A'}^2}\right) J^\beta  \,,
\end{align}
where  $P=p_1+p_2$ and $J^\beta$ is the final-state electromagnetic current, which is conserved, i.e. $P_\beta J^\beta =0$. 
This fact allows us 
 to write the corresponding cross section in a simple form without specifying ${\cal F}$. This is particularly useful for hadronic final states, for which the current receives non-perturbative contributions.
 
The corresponding cross section can be obtained from 
\begin{eqnarray}
4 p_\chi^\text{cm} \sqrt{s}\, \sigma \left(\chi_1 \chi_2 \to {\cal F}\right)&=&(2\pi)^4  \int \overline{|{\cal M} \left(\chi_1 \chi_2 \to {\cal F}\right)|^2 } d \Phi^{\cal F}(P)\\
&=& \frac{(2\pi)^4 e^2 g_X^2 \epsilon^2}{\left( s-m_{A'}^2\right)^2} \left(p_1^{\alpha}p_2^{\alpha'}+p_2^{\alpha}p_1^{\alpha'}-\frac{1}{2}g^{
\alpha \alpha'} (P^2-\Delta^2)\right)  \int d \Phi^{\cal F} (P) J_\alpha J^*_{\alpha'}\nonumber\,,
\label{eq:app_sigma1}
\end{eqnarray}  
where $p_\chi^\text{cm}$ is the 3-momentum of either particle in the $\chi_1\chi_2$ centre-of-mass frame, while $d\Phi^{\cal F}(P)$ is the phase-space element
\begin{eqnarray}
d\Phi^{\cal F}(P) =\delta^{(4)}\left(P- \sum_{i \in {\cal F}} p_i \right) \prod_{i \in{ \cal F}} \frac{d^3 p_i}{(2\pi)^3 2 E_i} \,.
\end{eqnarray}
Let us note that $\int d \Phi^{\cal F} (P) J_\alpha J^*_{\alpha'}$ is a  Lorentz-invariant function of only $P$, that vanishes if it is contracted with $P^\alpha$. As a result
\begin{align}
\int d \Phi^{\cal F} (P) J_\alpha J^*_{\alpha'}= \frac{1}{3} \left(g_{\alpha, \alpha'} - P_\alpha P_{\alpha'}/P^2 \right) \int J_\mu J^{\mu*} d\Phi^{\cal F}(P)\,.
\end{align}
Plugging this into Eq.~\eqref{eq:app_sigma1}, we obtain
\begin{equation}
\hspace{-0.6cm}
4 p_\chi^\text{cm} \sqrt{s}\,\sigma \left(\chi_1 \chi_2 \to {\cal F}\right)= - \frac{(2\pi)^4 e^2 g_X^2 \epsilon^2(s-\Delta^2)\left(2s +(2m_{\chi_1}+\Delta)^2\right)}{6s \left(s-m_{A'}^2\right)^2}  \int J_\mu J^{\mu*} d\Phi^{\cal F}.
\label{eq:app_sigma2}
\end{equation}  
For leptonic final states this can be evaluated perturbatively. For instance, for muon pairs we have
\begin{equation}
\hspace{-0.6cm}
4 p_\chi^\text{cm} \sqrt{s}\,\sigma \left(\chi_1 \chi_2 \to {\cal \mu^+\mu^-}\right)=  \frac{ e^2 g_X^2 \epsilon^2(s-\Delta^2)\left(2s +(2m_{\chi_1}+\Delta)^2\right)(s+2m_\mu^2)}{12\pi s \left(s-m_{A'}^2\right)^2}  \sqrt{1-\frac{4m_{\mu}^2}{s}}.
\label{eq:app_sigma3}
\end{equation} 
Interestingly,  one can apply the same reasoning we have used so far but for  $e^+e^-$ annihilations, obtaining Eq.~\eqref{eq:app_sigma2} with $\Delta\to0$, $m_{\chi_1}, m_{\chi_2}\to m_e$, $\epsilon\to1$  and $g_X\to e$.  This allows us to calculate $\int J_\mu J^{\mu*} d\Phi^{\cal F}(P)\big|_{hadrons}$ from the ratio $R(s)\equiv \sigma(e^-e^+\to hadrons)/\sigma(e^+e^-\to \mu^+\mu^- )$ as 
\begin{align}
\label{eq:ratio00}
 \int J_\mu J^{\mu*} d\Phi^{\cal F}(P)\big|_{hadrons}  = R(s) \int J_\mu J^{\mu*} d\Phi^{\cal F}(P) \big|_{\mu^+\mu^-}\,.
\end{align}
Hence 
\begin{align}
\label{eq:ratio0}
\frac{\sigma\left(\chi_1\chi_2 \to hadrons\right)}{\sigma\left(\chi_1\chi_2 \to \mu^+ \mu^- \right)} 
= R(s)\,.
\end{align}

\begin{figure*}[t]
\centering
\includegraphics[width=0.75\textwidth]{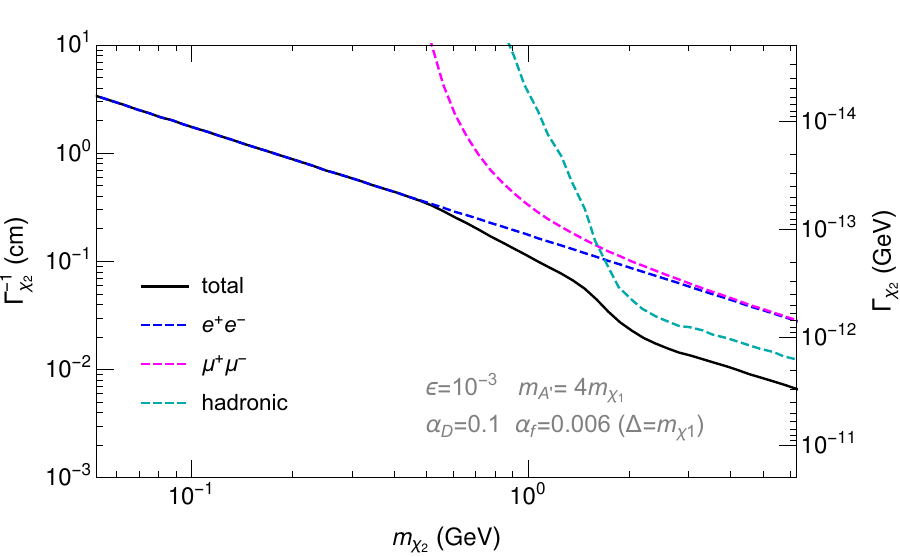}
\caption{ $\chi_2$ decay length together with the leptonic and the hadronic contributions for the parameters of Fig.~\ref{fig:africaANDrelic} (left).
}
\label{fig:chi2width}
\end{figure*}

\noindent \textbf{Decay width.} The decay rate of $\chi_2$ is given by

\begin{equation}
    \Gamma(\chi_2 \to \chi_1 {\cal F}) = \frac{(2\pi)^4}{2 m_{\chi_2}} \int \overline{|M(\chi_2 \to \chi_1 {\cal F})|^2} d\Phi^{{\cal F}+\chi_1} \left(P_{\chi_2}\right) \,.
\end{equation}
The crucial observation to relate this to our previous results is the fact that\footnote{The numerical factors in Eq.~\eqref{eq:Mref} are related to spin averaging.}
\begin{eqnarray}
2\overline{|M(\chi_2 \to \chi_1 {\cal F})|^2} &=& 4\overline{|M(\chi_1\chi_2 \to  {\cal F})|^2} \bigg|_{p_{\chi_1}\to-p_{\chi_1}}\,, \label{eq:Mref}\\
d\Phi^{{\cal F}+\chi_1} \left(P_{\chi_2}\right)  &=&  d\Phi^{{\cal F}} \left(P=P_{\chi_2}-P_{\chi_1}\right)\frac{d^3 p_{\chi_1}}{(2\pi)^3 2 E_{\chi_1}}\,.
\end{eqnarray}
Accordingly 
\begin{equation}
    \Gamma(\chi_2 \to \chi_1 {\cal F}) = \int \frac{d^3 p_{\chi_1}}{(2\pi)^3 2 E_{\chi_1}} \frac{\left[4 p_{\chi}^\text{cm} \sqrt{s} \sigma \left(\chi_1 \chi_2 \to {\cal F}\right)\right]_{s=(P_{\chi_2}-P_{\chi_1})^2}}{ m_{\chi_2}}    \,.
\end{equation}
In the $\chi_2$ rest frame,  $s=m_{\chi_1}^2+m_{\chi_2}^2- 2 m_{\chi_2} E_{\chi_1} = \Delta^2- 2m_{\chi_2}\left(E_{\chi_1} -m_{\chi_1}\right) \leq \Delta^2  $. In detail this implies 
\begin{equation}
\hspace{-0.3cm}
\frac{d^3 p_{\chi_1}}{(2\pi)^3 2 E_{\chi_1}} = \frac{4\pi |\vec{p}_{\chi_1}|^2 d |\vec{p}_{\chi_1}|}{(2\pi)^3 2 E_{\chi_1}} =\frac{4\pi |\vec{p}_{\chi_1}| E_{\chi_1} d E_{\chi_1} }{(2\pi)^3 2 E_{\chi_1}}  = \frac{4\pi |\vec{p}_{\chi_1}| E_{\chi_1} ds }{(2\pi)^3 (2 E_{\chi_1})(2m_{\chi_2})} = \frac{|\vec{p}_{\chi_1}| ds }{8\pi^2 m_{\chi_2}}\,,
\end{equation}
where 
\begin{eqnarray}
 | \vec{p}_{\chi_1}| =\sqrt{ 
 \left(
 \frac{s-m_{\chi_1}^2-m_{\chi_2}^2}{2m_{\chi_2}}
 \right)^2-m_{\chi_1}^2
 }\,.
 \label{eq:pchi1}
\end{eqnarray}
Thus
\begin{equation}
    \Gamma(\chi_2 \to \chi_1 {\cal F}) =  \frac{ 1  }{8\pi^2 m_{\chi_2}^2} \int^{\Delta^2}_{s_\text{min}}  ds | \vec{p}_{\chi_1}|   \left[4 p_{\chi}^\text{cm} \sqrt{s} \sigma \left(\chi_1 \chi_2 \to {\cal F}\right)\right]    \,,
\end{equation}
In particular, using Eqs.~\eqref{eq:app_sigma2} and \eqref{eq:ratio0} we conclude 
\begin{equation}
  \frac{ \Gamma(\chi_2 \to \chi_1 hadrons)}{\Gamma(\chi_2 \to \chi_1 \mu^+ \mu^-)} = \frac{\int^{\Delta^2}_{4m_\mu^2}  ds | \vec{p}_{\chi_1}|   \left[4 p_{\chi}^\text{cm} \sqrt{s} \sigma \left(\chi_1 \chi_2 \to  \mu^+\mu^-\right)\right]  R(s) }{\int^{\Delta^2}_{4m_\mu^2}  ds | \vec{p}_{\chi_1}|   \left[4 p_{\chi}^\text{cm} \sqrt{s} \sigma \left(\chi_1 \chi_2 \to  \mu^+\mu^-\right)\right]}  \,,
  \label{eq:Rsmaster}
\end{equation}
where the expression in the brackets is calculated using Eq.~\eqref{eq:app_sigma3}. In Fig.~\ref{fig:chi2width}, we illustrate this for the parameters of Fig.~\ref{fig:africaANDrelic} (left).

\bibliographystyle{JHEP_improved}
\bibliography{belle2_DarkHiggs_iDM}

\providecommand{\href}[2]{#2}\begingroup\raggedright\begin{thebibliography}{10}

\bibitem{Aprile:2018dbl}
{\bf XENON}, E.~Aprile et~al.,
  \href{http://dx.doi.org/10.1103/PhysRevLett.121.111302}{{\it {Dark Matter
  Search Results from a One Ton-Year Exposure of XENON1T}}, } {\em Phys. Rev.
  Lett.} {\bf 121} (2018), no.~11 111302,
  [\href{http://arxiv.org/abs/1805.12562}{{\tt 1805.12562}}].

\bibitem{Ren:2018gyx}
{\bf PandaX-II}, X.~Ren et~al.,
  \href{http://dx.doi.org/10.1103/PhysRevLett.121.021304}{{\it {Constraining
  Dark Matter Models with a Light Mediator at the PandaX-II Experiment}}, }
  {\em Phys. Rev. Lett.} {\bf 121} (2018), no.~2 021304,
  [\href{http://arxiv.org/abs/1802.06912}{{\tt 1802.06912}}].

\bibitem{Batell:2009di}
B.~Batell, M.~Pospelov, and A.~Ritz,
  \href{http://dx.doi.org/10.1103/PhysRevD.80.095024}{{\it {Exploring Portals
  to a Hidden Sector Through Fixed Targets}}, } {\em Phys. Rev. D} {\bf 80}
  (2009) 095024, [\href{http://arxiv.org/abs/0906.5614}{{\tt 0906.5614}}].

\bibitem{Batell:2009yf}
B.~Batell, M.~Pospelov, and A.~Ritz,
  \href{http://dx.doi.org/10.1103/PhysRevD.79.115008}{{\it {Probing a Secluded
  U(1) at B-factories}}, } {\em Phys. Rev.} {\bf D79} (2009) 115008,
  [\href{http://arxiv.org/abs/0903.0363}{{\tt 0903.0363}}].

\bibitem{Andreas:2012mt}
S.~Andreas, C.~Niebuhr, and A.~Ringwald,
  \href{http://dx.doi.org/10.1103/PhysRevD.86.095019}{{\it {New Limits on
  Hidden Photons from Past Electron Beam Dumps}}, } {\em Phys. Rev. D} {\bf 86}
  (2012) 095019, [\href{http://arxiv.org/abs/1209.6083}{{\tt 1209.6083}}].

\bibitem{Schmidt-Hoberg:2013hba}
K.~Schmidt-Hoberg, F.~Staub, and M.~W. Winkler,
  \href{http://dx.doi.org/10.1016/j.physletb.2013.11.015}{{\it {Constraints on
  light mediators: confronting dark matter searches with B physics}}, } {\em
  Phys. Lett. B} {\bf 727} (2013) 506--510,
  [\href{http://arxiv.org/abs/1310.6752}{{\tt 1310.6752}}].

\bibitem{Essig:2013vha}
R.~Essig, J.~Mardon, M.~Papucci, T.~Volansky, and Y.-M. Zhong,
  \href{http://dx.doi.org/10.1007/JHEP11(2013)167}{{\it {Constraining Light
  Dark Matter with Low-Energy $e^+e^-$ Colliders}}, } {\em JHEP} {\bf 11}
  (2013) 167, [\href{http://arxiv.org/abs/1309.5084}{{\tt 1309.5084}}].

\bibitem{Izaguirre:2013uxa}
E.~Izaguirre, G.~Krnjaic, P.~Schuster, and N.~Toro,
  \href{http://dx.doi.org/10.1103/PhysRevD.88.114015}{{\it {New Electron
  Beam-Dump Experiments to Search for MeV to few-GeV Dark Matter}}, } {\em
  Phys. Rev. D} {\bf 88} (2013) 114015,
  [\href{http://arxiv.org/abs/1307.6554}{{\tt 1307.6554}}].

\bibitem{Morrissey:2014yma}
D.~E. Morrissey and A.~P. Spray,
  \href{http://dx.doi.org/10.1007/JHEP06(2014)083}{{\it {New Limits on Light
  Hidden Sectors from Fixed-Target Experiments}}, } {\em JHEP} {\bf 06} (2014)
  083, [\href{http://arxiv.org/abs/1402.4817}{{\tt 1402.4817}}].

\bibitem{Batell:2014mga}
B.~Batell, R.~Essig, and Z.~Surujon,
  \href{http://dx.doi.org/10.1103/PhysRevLett.113.171802}{{\it {Strong
  Constraints on Sub-GeV Dark Sectors from SLAC Beam Dump E137}}, } {\em Phys.
  Rev. Lett.} {\bf 113} (2014), no.~17 171802,
  [\href{http://arxiv.org/abs/1406.2698}{{\tt 1406.2698}}].

\bibitem{Dolan:2014ska}
M.~J. Dolan, F.~Kahlhoefer, C.~McCabe, and K.~Schmidt-Hoberg,
  \href{http://dx.doi.org/10.1007/JHEP03(2015)171}{{\it {A taste of dark
  matter: Flavour constraints on pseudoscalar mediators}}, } {\em JHEP} {\bf
  03} (2015) 171, [\href{http://arxiv.org/abs/1412.5174}{{\tt 1412.5174}}].
  [Erratum: JHEP 07, 103 (2015)].

\bibitem{Krnjaic:2015mbs}
G.~Krnjaic, \href{http://dx.doi.org/10.1103/PhysRevD.94.073009}{{\it {Probing
  Light Thermal Dark-Matter With a Higgs Portal Mediator}}, } {\em Phys. Rev.
  D} {\bf 94} (2016), no.~7 073009,
  [\href{http://arxiv.org/abs/1512.04119}{{\tt 1512.04119}}].

\bibitem{Dolan:2017osp}
M.~J. Dolan, T.~Ferber, C.~Hearty, F.~Kahlhoefer, and K.~Schmidt-Hoberg,
  \href{http://dx.doi.org/10.1007/JHEP12(2017)094}{{\it {Revised constraints
  and Belle II sensitivity for visible and invisible axion-like particles}}, }
  {\em JHEP} {\bf 12} (2017) 094, [\href{http://arxiv.org/abs/1709.00009}{{\tt
  1709.00009}}].

\bibitem{Knapen:2017xzo}
S.~Knapen, T.~Lin, and K.~M. Zurek,
  \href{http://dx.doi.org/10.1103/PhysRevD.96.115021}{{\it {Light Dark Matter:
  Models and Constraints}}, } {\em Phys. Rev. D} {\bf 96} (2017), no.~11
  115021, [\href{http://arxiv.org/abs/1709.07882}{{\tt 1709.07882}}].

\bibitem{Beacham:2019nyx}
J.~Beacham et~al., \href{http://dx.doi.org/10.1088/1361-6471/ab4cd2}{{\it
  {Physics Beyond Colliders at CERN: Beyond the Standard Model Working Group
  Report}}, } {\em J. Phys. G} {\bf 47} (2020), no.~1 010501,
  [\href{http://arxiv.org/abs/1901.09966}{{\tt 1901.09966}}].

\bibitem{Bernreuther:2019pfb}
E.~Bernreuther, F.~Kahlhoefer, M.~Kr\"amer, and P.~Tunney,
  \href{http://dx.doi.org/10.1007/JHEP01(2020)162}{{\it {Strongly interacting
  dark sectors in the early Universe and at the LHC through a simplified
  portal}}, } {\em JHEP} {\bf 01} (2020) 162,
  [\href{http://arxiv.org/abs/1907.04346}{{\tt 1907.04346}}].

\bibitem{Bondarenko:2019vrb}
K.~Bondarenko, A.~Boyarsky, T.~Bringmann, M.~Hufnagel, K.~Schmidt-Hoberg,
  et~al., \href{http://dx.doi.org/10.1007/JHEP03(2020)118}{{\it {Direct
  detection and complementary constraints for sub-GeV dark matter}}, } {\em
  JHEP} {\bf 03} (2020) 118, [\href{http://arxiv.org/abs/1909.08632}{{\tt
  1909.08632}}].

\bibitem{Filimonova:2019tuy}
A.~Filimonova, R.~Sch\"afer, and S.~Westhoff,
  \href{http://dx.doi.org/10.1103/PhysRevD.101.095006}{{\it {Probing dark
  sectors with long-lived particles at BELLE II}}, } {\em Phys. Rev. D} {\bf
  101} (2020), no.~9 095006, [\href{http://arxiv.org/abs/1911.03490}{{\tt
  1911.03490}}].

\bibitem{Ballett:2019pyw}
P.~Ballett, M.~Hostert, and S.~Pascoli,
  \href{http://dx.doi.org/10.1103/PhysRevD.101.115025}{{\it {Dark Neutrinos and
  a Three Portal Connection to the Standard Model}}, } {\em Phys. Rev. D} {\bf
  101} (2020), no.~11 115025, [\href{http://arxiv.org/abs/1903.07589}{{\tt
  1903.07589}}].

\bibitem{BelleII:2020fag}
{\bf Belle-II}, F.~Abudin\'en et~al.,
  \href{http://dx.doi.org/10.1103/PhysRevLett.125.161806}{{\it {Search for
  Axion-Like Particles produced in $e^+e^-$ collisions at Belle II}}, } {\em
  Phys. Rev. Lett.} {\bf 125} (2020), no.~16 161806,
  [\href{http://arxiv.org/abs/2007.13071}{{\tt 2007.13071}}].

\bibitem{Bernal:2017mqb}
N.~Bernal, X.~Chu, and J.~Pradler,
  \href{http://dx.doi.org/10.1103/PhysRevD.95.115023}{{\it {Simply split
  strongly interacting massive particles}}, } {\em Phys. Rev. D} {\bf 95}
  (2017), no.~11 115023, [\href{http://arxiv.org/abs/1702.04906}{{\tt
  1702.04906}}].

\bibitem{Jodlowski:2019ycu}
K.~Jod\l{}owski, F.~Kling, L.~Roszkowski, and S.~Trojanowski,
  \href{http://dx.doi.org/10.1103/PhysRevD.101.095020}{{\it {Extending the
  reach of FASER, MATHUSLA, and SHiP towards smaller lifetimes using secondary
  particle production}}, } {\em Phys. Rev. D} {\bf 101} (2020), no.~9 095020,
  [\href{http://arxiv.org/abs/1911.11346}{{\tt 1911.11346}}].

\bibitem{Baek:2020owl}
S.~Baek, J.~Kim, and P.~Ko,
  \href{http://dx.doi.org/10.1016/j.physletb.2020.135848}{{\it {XENON1T excess
  in local $Z_2$ DM models with light dark sector}}, } {\em Phys. Lett. B} {\bf
  810} (2020) 135848, [\href{http://arxiv.org/abs/2006.16876}{{\tt
  2006.16876}}].

\bibitem{Hostert:2020xku}
M.~Hostert and M.~Pospelov, {\it {Novel multi-lepton signatures of dark sectors
  in light meson decays}},  \href{http://arxiv.org/abs/2012.02142}{{\tt
  2012.02142}}.

\bibitem{TuckerSmith:2001hy}
D.~Tucker-Smith and N.~Weiner,
  \href{http://dx.doi.org/10.1103/PhysRevD.64.043502}{{\it {Inelastic dark
  matter}}, } {\em Phys. Rev. D} {\bf 64} (2001) 043502,
  [\href{http://arxiv.org/abs/hep-ph/0101138}{{\tt hep-ph/0101138}}].

\bibitem{Bernreuther:2020koj}
E.~Bernreuther, S.~Heeba, and F.~Kahlhoefer, {\it {Resonant Sub-GeV Dirac Dark
  Matter}},  \href{http://arxiv.org/abs/2010.14522}{{\tt 2010.14522}}.

\bibitem{Depta:2019lbe}
P.~F. Depta, M.~Hufnagel, K.~Schmidt-Hoberg, and S.~Wild,
  \href{http://dx.doi.org/10.1088/1475-7516/2019/04/029}{{\it {BBN constraints
  on the annihilation of MeV-scale dark matter}}, } {\em JCAP} {\bf 04} (2019)
  029, [\href{http://arxiv.org/abs/1901.06944}{{\tt 1901.06944}}].

\bibitem{Izaguirre:2015zva}
E.~Izaguirre, G.~Krnjaic, and B.~Shuve,
  \href{http://dx.doi.org/10.1103/PhysRevD.93.063523}{{\it {Discovering
  Inelastic Thermal-Relic Dark Matter at Colliders}}, } {\em Phys. Rev. D} {\bf
  93} (2016), no.~6 063523, [\href{http://arxiv.org/abs/1508.03050}{{\tt
  1508.03050}}].

\bibitem{Izaguirre:2017bqb}
E.~Izaguirre, Y.~Kahn, G.~Krnjaic, and M.~Moschella,
  \href{http://dx.doi.org/10.1103/PhysRevD.96.055007}{{\it {Testing Light Dark
  Matter Coannihilation With Fixed-Target Experiments}}, } {\em Phys. Rev. D}
  {\bf 96} (2017), no.~5 055007, [\href{http://arxiv.org/abs/1703.06881}{{\tt
  1703.06881}}].

\bibitem{Berlin:2018jbm}
A.~Berlin and F.~Kling,
  \href{http://dx.doi.org/10.1103/PhysRevD.99.015021}{{\it {Inelastic Dark
  Matter at the LHC Lifetime Frontier: ATLAS, CMS, LHCb, CODEX-b, FASER, and
  MATHUSLA}}, } {\em Phys. Rev. D} {\bf 99} (2019), no.~1 015021,
  [\href{http://arxiv.org/abs/1810.01879}{{\tt 1810.01879}}].

\bibitem{Duerr:2019dmv}
M.~Duerr, T.~Ferber, C.~Hearty, F.~Kahlhoefer, K.~Schmidt-Hoberg, et~al.,
  \href{http://dx.doi.org/10.1007/JHEP02(2020)039}{{\it {Invisible and
  displaced dark matter signatures at Belle II}}, } {\em JHEP} {\bf 02} (2020)
  039, [\href{http://arxiv.org/abs/1911.03176}{{\tt 1911.03176}}].

\bibitem{Kahlhoefer:2015bea}
F.~Kahlhoefer, K.~Schmidt-Hoberg, T.~Schwetz, and S.~Vogl,
  \href{http://dx.doi.org/10.1007/JHEP02(2016)016}{{\it {Implications of
  unitarity and gauge invariance for simplified dark matter models}}, } {\em
  JHEP} {\bf 02} (2016) 016, [\href{http://arxiv.org/abs/1510.02110}{{\tt
  1510.02110}}].

\bibitem{Duerr:2017uap}
M.~Duerr, A.~Grohsjean, F.~Kahlhoefer, B.~Penning, K.~Schmidt-Hoberg, et~al.,
  \href{http://dx.doi.org/10.1007/JHEP04(2017)143}{{\it {Hunting the dark
  Higgs}}, } {\em JHEP} {\bf 04} (2017) 143,
  [\href{http://arxiv.org/abs/1701.08780}{{\tt 1701.08780}}].

\bibitem{Darme:2018jmx}
L.~Darm\'e, S.~Rao, and L.~Roszkowski,
  \href{http://dx.doi.org/10.1007/JHEP12(2018)014}{{\it {Signatures of dark
  Higgs boson in light fermionic dark matter scenarios}}, } {\em JHEP} {\bf 12}
  (2018) 014, [\href{http://arxiv.org/abs/1807.10314}{{\tt 1807.10314}}].

\bibitem{TheBelle:2015mwa}
{\bf Belle}, I.~Jaegle,
  \href{http://dx.doi.org/10.1103/PhysRevLett.114.211801}{{\it {Search for the
  dark photon and the dark Higgs boson at Belle}}, } {\em Phys. Rev. Lett.}
  {\bf 114} (2015), no.~21 211801, [\href{http://arxiv.org/abs/1502.00084}{{\tt
  1502.00084}}].

\bibitem{Babu:1997st}
K.~S. Babu, C.~F. Kolda, and J.~March-Russell,
  \href{http://dx.doi.org/10.1103/PhysRevD.57.6788}{{\it {Implications of
  generalized Z - Z-prime mixing}}, } {\em Phys. Rev.} {\bf D57} (1998)
  6788--6792, [\href{http://arxiv.org/abs/hep-ph/9710441}{{\tt
  hep-ph/9710441}}].

\bibitem{Frandsen:2011cg}
M.~T. Frandsen, F.~Kahlhoefer, S.~Sarkar, and K.~Schmidt-Hoberg,
  \href{http://dx.doi.org/10.1007/JHEP09(2011)128}{{\it {Direct detection of
  dark matter in models with a light Z'}}, } {\em JHEP} {\bf 09} (2011) 128,
  [\href{http://arxiv.org/abs/1107.2118}{{\tt 1107.2118}}].

\bibitem{Duerr:2016tmh}
M.~Duerr, F.~Kahlhoefer, K.~Schmidt-Hoberg, T.~Schwetz, and S.~Vogl,
  \href{http://dx.doi.org/10.1007/JHEP09(2016)042}{{\it {How to save the WIMP:
  global analysis of a dark matter model with two s-channel mediators}}, } {\em
  JHEP} {\bf 09} (2016) 042, [\href{http://arxiv.org/abs/1606.07609}{{\tt
  1606.07609}}].

\bibitem{Darme:2017glc}
L.~Darm\'e, S.~Rao, and L.~Roszkowski,
  \href{http://dx.doi.org/10.1007/JHEP03(2018)084}{{\it {Light dark Higgs boson
  in minimal sub-GeV dark matter scenarios}}, } {\em JHEP} {\bf 03} (2018) 084,
  [\href{http://arxiv.org/abs/1710.08430}{{\tt 1710.08430}}].

\bibitem{Tanabashi:2018oca}
{\bf Particle Data Group}, M.~Tanabashi et~al.,
  \href{http://dx.doi.org/10.1103/PhysRevD.98.030001}{{\it {Review of Particle
  Physics}}, } {\em Phys. Rev. D} {\bf 98} (2018), no.~3 030001.

\bibitem{Ko:2019wxq}
P.~Ko, T.~Matsui, and Y.-L. Tang, {\it {Dark Matter Bound State Formation in
  Fermionic $Z_2$ DM model with Light Dark Photon and Dark Higgs Boson}},
  \href{http://arxiv.org/abs/1910.04311}{{\tt 1910.04311}}.

\bibitem{Bringmann:2020mgx}
T.~Bringmann, P.~F. Depta, M.~Hufnagel, and K.~Schmidt-Hoberg, {\it {Precise
  dark matter relic abundance in decoupled sectors}},
  \href{http://arxiv.org/abs/2007.03696}{{\tt 2007.03696}}.

\bibitem{Belanger:2018ccd}
G.~B\'elanger, F.~Boudjema, A.~Goudelis, A.~Pukhov, and B.~Zaldivar,
  \href{http://dx.doi.org/10.1016/j.cpc.2018.04.027}{{\it {micrOMEGAs5.0 :
  Freeze-in}}, } {\em Comput. Phys. Commun.} {\bf 231} (2018) 173--186,
  [\href{http://arxiv.org/abs/1801.03509}{{\tt 1801.03509}}].

\bibitem{Ade:2015xua}
{\bf Planck}, P.~Ade et~al.,
  \href{http://dx.doi.org/10.1051/0004-6361/201525830}{{\it {Planck 2015
  results. XIII. Cosmological parameters}}, } {\em Astron. Astrophys.} {\bf
  594} (2016) A13, [\href{http://arxiv.org/abs/1502.01589}{{\tt 1502.01589}}].

\bibitem{An:2016kie}
H.~An, M.~B. Wise, and Y.~Zhang,
  \href{http://dx.doi.org/10.1016/j.physletb.2017.08.010}{{\it {Strong CMB
  Constraint On P-Wave Annihilating Dark Matter}}, } {\em Phys. Lett. B} {\bf
  773} (2017) 121--124, [\href{http://arxiv.org/abs/1606.02305}{{\tt
  1606.02305}}].

\bibitem{DAgnolo:2015ujb}
R.~T. D'Agnolo and J.~T. Ruderman,
  \href{http://dx.doi.org/10.1103/PhysRevLett.115.061301}{{\it {Light Dark
  Matter from Forbidden Channels}}, } {\em Phys. Rev. Lett.} {\bf 115} (2015),
  no.~6 061301, [\href{http://arxiv.org/abs/1505.07107}{{\tt 1505.07107}}].

\bibitem{Hook:2010tw}
A.~Hook, E.~Izaguirre, and J.~G. Wacker,
  \href{http://dx.doi.org/10.1155/2011/859762}{{\it {Model Independent Bounds
  on Kinetic Mixing}}, } {\em Adv. High Energy Phys.} {\bf 2011} (2011) 859762,
  [\href{http://arxiv.org/abs/1006.0973}{{\tt 1006.0973}}].

\bibitem{Kribs:2020vyk}
G.~D. Kribs, D.~McKeen, and N.~Raj, {\it {Breaking up the Proton: An Affair
  with Dark Forces}},  \href{http://arxiv.org/abs/2007.15655}{{\tt
  2007.15655}}.

\bibitem{Winkler:2018qyg}
M.~W. Winkler, \href{http://dx.doi.org/10.1103/PhysRevD.99.015018}{{\it {Decay
  and detection of a light scalar boson mixing with the Higgs boson}}, } {\em
  Phys. Rev. D} {\bf 99} (2019), no.~1 015018,
  [\href{http://arxiv.org/abs/1809.01876}{{\tt 1809.01876}}].

\bibitem{Berger:2016vxi}
J.~Berger, K.~Jedamzik, and D.~G.~E. Walker,
  \href{http://dx.doi.org/10.1088/1475-7516/2016/11/032}{{\it {Cosmological
  Constraints on Decoupled Dark Photons and Dark Higgs}}, } {\em JCAP} {\bf 11}
  (2016) 032, [\href{http://arxiv.org/abs/1605.07195}{{\tt 1605.07195}}].

\bibitem{Fradette:2018hhl}
A.~Fradette, M.~Pospelov, J.~Pradler, and A.~Ritz,
  \href{http://dx.doi.org/10.1103/PhysRevD.99.075004}{{\it {Cosmological beam
  dump: constraints on dark scalars mixed with the Higgs boson}}, } {\em Phys.
  Rev. D} {\bf 99} (2019), no.~7 075004,
  [\href{http://arxiv.org/abs/1812.07585}{{\tt 1812.07585}}].

\bibitem{Depta:2020zbh}
P.~F. Depta, M.~Hufnagel, and K.~Schmidt-Hoberg, {\it {Updated BBN constraints
  on electromagnetic decays of MeV-scale particles}},
  \href{http://arxiv.org/abs/2011.06519}{{\tt 2011.06519}}.

\bibitem{Bar:2019ifz}
N.~Bar, K.~Blum, and G.~D'Amico,
  \href{http://dx.doi.org/10.1103/PhysRevD.101.123025}{{\it {Is there a
  supernova bound on axions?}}, } {\em Phys. Rev. D} {\bf 101} (2020), no.~12
  123025, [\href{http://arxiv.org/abs/1907.05020}{{\tt 1907.05020}}].

\bibitem{deNiverville:2011it}
P.~deNiverville, M.~Pospelov, and A.~Ritz,
  \href{http://dx.doi.org/10.1103/PhysRevD.84.075020}{{\it {Observing a light
  dark matter beam with neutrino experiments}}, } {\em Phys. Rev. D} {\bf 84}
  (2011) 075020, [\href{http://arxiv.org/abs/1107.4580}{{\tt 1107.4580}}].

\bibitem{Bjorken:1988as}
J.~Bjorken, S.~Ecklund, W.~Nelson, A.~Abashian, C.~Church, et~al.,
  \href{http://dx.doi.org/10.1103/PhysRevD.38.3375}{{\it {Search for Neutral
  Metastable Penetrating Particles Produced in the SLAC Beam Dump}}, } {\em
  Phys. Rev. D} {\bf 38} (1988) 3375.

\bibitem{Berlin:2018pwi}
A.~Berlin, S.~Gori, P.~Schuster, and N.~Toro,
  \href{http://dx.doi.org/10.1103/PhysRevD.98.035011}{{\it {Dark Sectors at the
  Fermilab SeaQuest Experiment}}, } {\em Phys. Rev. D} {\bf 98} (2018), no.~3
  035011, [\href{http://arxiv.org/abs/1804.00661}{{\tt 1804.00661}}].

\bibitem{Aguilar-Arevalo:2017mqx}
{\bf MiniBooNE}, A.~Aguilar-Arevalo et~al.,
  \href{http://dx.doi.org/10.1103/PhysRevLett.118.221803}{{\it {Dark Matter
  Search in a Proton Beam Dump with MiniBooNE}}, } {\em Phys. Rev. Lett.} {\bf
  118} (2017), no.~22 221803, [\href{http://arxiv.org/abs/1702.02688}{{\tt
  1702.02688}}].

\bibitem{NA64:2019imj}
D.~Banerjee et~al.,
  \href{http://dx.doi.org/10.1103/PhysRevLett.123.121801}{{\it {Dark matter
  search in missing energy events with NA64}}, } {\em Phys. Rev. Lett.} {\bf
  123} (2019), no.~12 121801, [\href{http://arxiv.org/abs/1906.00176}{{\tt
  1906.00176}}].

\bibitem{Lees:2017lec}
{\bf BaBar}, J.~Lees et~al.,
  \href{http://dx.doi.org/10.1103/PhysRevLett.119.131804}{{\it {Search for
  Invisible Decays of a Dark Photon Produced in ${e}^{+}{e}^{-}$ Collisions at
  BaBar}}, } {\em Phys. Rev. Lett.} {\bf 119} (2017), no.~13 131804,
  [\href{http://arxiv.org/abs/1702.03327}{{\tt 1702.03327}}].

\bibitem{Feng:2017uoz}
J.~L. Feng, I.~Galon, F.~Kling, and S.~Trojanowski,
  \href{http://dx.doi.org/10.1103/PhysRevD.97.035001}{{\it {ForwArd Search
  ExpeRiment at the LHC}}, } {\em Phys. Rev. D} {\bf 97} (2018), no.~3 035001,
  [\href{http://arxiv.org/abs/1708.09389}{{\tt 1708.09389}}].

\bibitem{Chou:2016lxi}
J.~P. Chou, D.~Curtin, and H.~Lubatti,
  \href{http://dx.doi.org/10.1016/j.physletb.2017.01.043}{{\it {New Detectors
  to Explore the Lifetime Frontier}}, } {\em Phys. Lett. B} {\bf 767} (2017)
  29--36, [\href{http://arxiv.org/abs/1606.06298}{{\tt 1606.06298}}].

\bibitem{Gligorov:2017nwh}
V.~V. Gligorov, S.~Knapen, M.~Papucci, and D.~J. Robinson,
  \href{http://dx.doi.org/10.1103/PhysRevD.97.015023}{{\it {Searching for
  Long-lived Particles: A Compact Detector for Exotics at LHCb}}, } {\em Phys.
  Rev. D} {\bf 97} (2018), no.~1 015023,
  [\href{http://arxiv.org/abs/1708.09395}{{\tt 1708.09395}}].

\bibitem{Akesson:2018vlm}
{\bf LDMX}, T.~\r{A}kesson et~al., {\it {Light Dark Matter eXperiment (LDMX)}},
   \href{http://arxiv.org/abs/1808.05219}{{\tt 1808.05219}}.

\bibitem{Abe:2010gxa}
{\bf Belle-II}, T.~Abe et~al., {\it {Belle II Technical Design Report}},
  \href{http://arxiv.org/abs/1011.0352}{{\tt 1011.0352}}.

\bibitem{Alloul:2013bka}
A.~Alloul, N.~D. Christensen, C.~Degrande, C.~Duhr, and B.~Fuks,
  \href{http://dx.doi.org/10.1016/j.cpc.2014.04.012}{{\it {FeynRules 2.0 - A
  complete toolbox for tree-level phenomenology}}, } {\em Comput. Phys.
  Commun.} {\bf 185} (2014) 2250--2300,
  [\href{http://arxiv.org/abs/1310.1921}{{\tt 1310.1921}}].

\bibitem{Degrande:2011ua}
C.~Degrande, C.~Duhr, B.~Fuks, D.~Grellscheid, O.~Mattelaer, et~al.,
  \href{http://dx.doi.org/10.1016/j.cpc.2012.01.022}{{\it {UFO - The Universal
  FeynRules Output}}, } {\em Comput. Phys. Commun.} {\bf 183} (2012)
  1201--1214, [\href{http://arxiv.org/abs/1108.2040}{{\tt 1108.2040}}].

\bibitem{Alwall:2014hca}
J.~Alwall, R.~Frederix, S.~Frixione, V.~Hirschi, F.~Maltoni, et~al.,
  \href{http://dx.doi.org/10.1007/JHEP07(2014)079}{{\it {The automated
  computation of tree-level and next-to-leading order differential cross
  sections, and their matching to parton shower simulations}}, } {\em JHEP}
  {\bf 07} (2014) 079, [\href{http://arxiv.org/abs/1405.0301}{{\tt
  1405.0301}}].

\bibitem{Ilten:2018crw}
P.~Ilten, Y.~Soreq, M.~Williams, and W.~Xue,
  \href{http://dx.doi.org/10.1007/JHEP06(2018)004}{{\it {Serendipity in dark
  photon searches}}, } {\em JHEP} {\bf 06} (2018) 004,
  [\href{http://arxiv.org/abs/1801.04847}{{\tt 1801.04847}}].

\bibitem{Liu:2014cma}
J.~Liu, N.~Weiner, and W.~Xue,
  \href{http://dx.doi.org/10.1007/JHEP08(2015)050}{{\it {Signals of a Light
  Dark Force in the Galactic Center}}, } {\em JHEP} {\bf 08} (2015) 050,
  [\href{http://arxiv.org/abs/1412.1485}{{\tt 1412.1485}}].

\bibitem{Kou:2018nap}
{\bf Belle-II}, W.~Altmannshofer et~al.,
  \href{http://dx.doi.org/10.1093/ptep/ptz106}{{\it {The Belle II Physics
  Book}}, } {\em PTEP} {\bf 2019} (2019), no.~12 123C01,
  [\href{http://arxiv.org/abs/1808.10567}{{\tt 1808.10567}}]. [Erratum: PTEP
  2020, 029201 (2020)].

\bibitem{Pospelov:2008zw}
M.~Pospelov, \href{http://dx.doi.org/10.1103/PhysRevD.80.095002}{{\it {Secluded
  U(1) below the weak scale}}, } {\em Phys. Rev. D} {\bf 80} (2009) 095002,
  [\href{http://arxiv.org/abs/0811.1030}{{\tt 0811.1030}}].

\bibitem{Mohlabeng:2019vrz}
G.~Mohlabeng, \href{http://dx.doi.org/10.1103/PhysRevD.99.115001}{{\it
  {Revisiting the dark photon explanation of the muon anomalous magnetic
  moment}}, } {\em Phys. Rev. D} {\bf 99} (2019), no.~11 115001,
  [\href{http://arxiv.org/abs/1902.05075}{{\tt 1902.05075}}].

\bibitem{Abi:2021gix}
{\bf Muon g-2}, B.~Abi et~al.,
  \href{http://dx.doi.org/10.1103/PhysRevLett.126.141801}{{\it {Measurement of
  the Positive Muon Anomalous Magnetic Moment to 0.46 ppm}}, } {\em Phys. Rev.
  Lett.} {\bf 126} (2021) 141801, [\href{http://arxiv.org/abs/2104.03281}{{\tt
  2104.03281}}].

\bibitem{Chen:2015vqy}
C.-Y. Chen, H.~Davoudiasl, W.~J. Marciano, and C.~Zhang,
  \href{http://dx.doi.org/10.1103/PhysRevD.93.035006}{{\it {Implications of a
  light \textquotedblleft{}dark Higgs\textquotedblright{} solution to the
  $g_?$-2 discrepancy}}, } {\em Phys. Rev. D} {\bf 93} (2016), no.~3 035006,
  [\href{http://arxiv.org/abs/1511.04715}{{\tt 1511.04715}}].

\bibitem{Agnes:2018ves}
{\bf DarkSide}, P.~Agnes et~al.,
  \href{http://dx.doi.org/10.1103/PhysRevLett.121.081307}{{\it {Low-Mass Dark
  Matter Search with the DarkSide-50 Experiment}}, } {\em Phys. Rev. Lett.}
  {\bf 121} (2018), no.~8 081307, [\href{http://arxiv.org/abs/1802.06994}{{\tt
  1802.06994}}].

\bibitem{Ariga:2018uku}
{\bf FASER}, A.~Ariga et~al.,
  \href{http://dx.doi.org/10.1103/PhysRevD.99.095011}{{\it
  {FASER\textquoteright{}s physics reach for long-lived particles}}, } {\em
  Phys. Rev. D} {\bf 99} (2019), no.~9 095011,
  [\href{http://arxiv.org/abs/1811.12522}{{\tt 1811.12522}}].

\bibitem{Bennett:2006fi}
{\bf Muon g-2}, G.~W. Bennett et~al.,
  \href{http://dx.doi.org/10.1103/PhysRevD.73.072003}{{\it {Final Report of the
  Muon E821 Anomalous Magnetic Moment Measurement at BNL}}, } {\em Phys. Rev.
  D} {\bf 73} (2006) 072003, [\href{http://arxiv.org/abs/hep-ex/0602035}{{\tt
  hep-ex/0602035}}].

\bibitem{Borsanyi:2020mff}
S.~Borsanyi et~al., {\it {Leading hadronic contribution to the muon 2 magnetic
  moment from lattice QCD}},  \href{http://arxiv.org/abs/2002.12347}{{\tt
  2002.12347}}.

\end{thebibliography}\endgroup

\end{document}